\documentclass{article}
\usepackage{a4,amsmath,epsfig,amssymb}
\newcommand{\bm}[1]{\boldsymbol #1}
\newcommand{\zmpf}[1]{\mbox{\hspace{#1em}}}
\newcommand{\Id}{\mbox{$\,$\rm 1\zmpf{-0.62}{\small 1}}}
\newcommand{\RR}{\mathbb R}

\begin{document}

\title{Four-state quantum chain\\ as a model of sequence evolution}
\author{{\sc Joachim Hermisson$^{1,2}$, Holger Wagner$^{3}$ and 
Michael Baake$^{1}$} 
\\[2mm]
${}^{1}$Institut f\"ur Theoretische Physik, Universit\"at
T\"ubingen,\\ Auf der Morgenstelle 14, 72076 T\"ubingen, Germany\\
${}^{2}$Institut f\"ur Theorie der Kondensierten Materie,\\ 
Universit\"at Karlsruhe, 76128 Karlsruhe, Germany\\
${}^{3}$Max-Planck-Institut f\"ur Biophysikalische Chemie,\\
Am Fa{\ss}berg 11, 37077 G\"ottingen, Germany}
\maketitle
\begin{abstract}
A variety of selection-mutation models for DNA (or RNA) sequences, 
well known in molecular evolution, can be translated into a model of coupled
Ising quantum chains. This correspondence is used to investigate the 
genetic variability and error threshold behaviour in dependence of possible 
fitness landscapes. In contrast to the two-state models treated
hitherto, the model explicitly takes the four-state nature of the
nucleotide alphabet into account and allowes for the distinction of
mutation rates for the different base substitutions, as given by
standard mutation schemes of molecular phylogeny. As a consequence of
this refined treatment, new phase diagrams for the error threshold
behaviour are obtained, with appearance of a novel phase in which the
nucleotide ordering of the wildtype sequence is only partially conserved.
Explicit analytic and numeric results are presented for evolution
dynamics and equilibrium behaviour in a number of accessible
situations, such as quadratic fitness landscapes and the Kimura 
2 parameter mutation scheme. 
\end{abstract}

\section{Introduction}

One prominent phenomenon in the theory of molecular evolution that has 
also attracted considerable attention in statistical physics is the 
so-called {\em error threshold}. It describes the breakdown of 
genetic order in mutation-selection models for mutation rates 
surpassing a certain critical value. The prototype model for the
description of the error threshold is Eigen's quasispecies model 
in sequence space \cite{E,ECS} (which is effectively equivalent to a 
coupled mutation-selection model in population genetics, cf \cite{CK}),
originally designed for the description of prebiotic RNA
evolution. However, the threshold is supposed to be a phenomenon that 
should occur in a rather general class of mutation-selection models.

In order to set up a mutation-selection model that is tractable by 
analytical (or at least numerical) methods, severe simplifications 
of the original biological situation seem to be indispensable. 
Analytical approaches generally have to restrict to the treatment 
of infinitly large populations and rather simple fitness functions,
such as the sharply peaked landscape of Eigens original model.
Another common approximation, also used in previous studies of the 
quasispecies model, amounts for the simplified 
representation of genotypes as binary strings. In the context of
molecular evolutionary theory, this may be thought of as representing 
DNA or RNA strands by sequences of {\em purins} and {\em pyrimidins},
hence with only two states per site, neglecting the fact that genetic 
information is really given by a four-letter alphabet. In this
article, we present a four-state mutation-selection model
which is capable to describe the full nucleotide alphabet and
incorporates the standard mutation schemes of molecular phylogeny. 
In particular, the phase diagramms are discussed in detail which 
are more polymorphic than for the two-state model. This shows that, 
for a full understanding of the error threshold behaviour in 
molecular evolution, investigations can not be restricted entirely 
to the study of two-state models.

One important step towards an understanding of the
threshold phenomenon has been its identification with an equilibrium phase
transition in physics by the translation of a time-discrete version of
the quasispecies model into the transfer matrix of an anisotropic 
two-dimensional Ising model \cite{Leut}. This equivalence was further
exploited to study various aspects of the error threshold
with methods from statistical physics \cite{Leut2,Tara,FPS,FP,MT}. 
It turns out, however, that the anisotropy of that model is not so
easy to handle and the analysis of the relevant biological quantities 
(which correspond to certain surface properties of the Ising model)
remains an involved problem. Due to the complications of the model,
almost all results obtained so far are approximate or numerical. The 
only exact result for the {\em sharply peaked landscape} \cite{Gal}
has been worked out via a different analogy to a model of directed
polymers, using the specific properties of that very special fitness 
landscape.

An alternative approach to the analysis of mutation-selection models
and the error threshold which avoids some of the problems of the
anisotropic Ising model has been brought up in \cite{BBW,WBG}.   
Here, the starting point on the biological side is a slightly changed 
model which describes the evolution of a population with overlapping 
generations in continuous time. It turns out that, after a
reformulation in tensor products, the two-state version of this model 
is equivalent to the Hamiltonian of an Ising quantum chain. Thereby, 
the change to continuous time in the biological description
corresponds to the anisotropic limit that connects the 
two-dimensional Ising model and the quantum chain in physics 
(cf.~\cite{Kogut}). The quantum chain model is technically easier to 
handle, and exact results for two non-trivial fitness landscapes, 
namely Onsager's landscape and the quadratic fitness function, have 
been worked out \cite{BBW,WBG}. 

Accordingly, we extend this latter approach to a full four-state model
in this study. The quantum chain analogy allows to use well-known methods 
from statistical mechanics for the solution of the model, so that we do not
have to dwell on technical details here. For an extended presentation 
of methods (with regard to the two-state model) using techniques from
rigorous mean field theory, we refer to \cite{Wag,WBG}. The main focus 
is instead on the discussion of the threshold behaviour and in
particular the increased complexity of the phase diagram due to the 
consideration of the four-state nature of biological information and 
the refined schemes of molecular mutation rates. 

In the following section, we start with a presentation of the biological 
foundations of our model. Only thereafter, we will introduce the quantum
chain model in Section 3. In Section 4, analytical and numerical
results are presented for a number of specific four-state models 
with permutation invariant fitness landscapes. Also the properties of
finite sequences and the evolution dynamics will be studied. 
We close with a summary of our results and a discussion of open
problems in Section 5.

\section{Biological foundations}

Genetic information is coded in DNA (and RNA) molecules. These are 
heteropolymers of four units (nucleotides) which differ in a specific
base. The essential aspect of a DNA sequence is captured in
a string over a four-letter alphabet
\begin{equation}
{\bm \sigma} \in V \equiv V_1 \times V_2 \times \dots \times V_N \;;\quad
V_i = \{A,C,G,T\}
\end{equation}
where each letter represents a particular base: $A$ and $G$  for
adenine and guanine (the purins), $C$ and $T$ for cytosine and thymine 
(the pyrimidins). In RNA sequences, $T$ is replaced by $U$ for uracil.
We will therefore treat the $4^N$ different sequences of a fixed, 
finite length $N$ as our genotypes (which may be thought of as coding
for something, such as a virus or an enzyme). Disregarding
environmental effects, we may identify a collection of genotypes with
a {\em population} of haploid `individuals'. Evolution then describes
the change of the population composition in time. 

A standard model for the evolution of an infinite, asexually 
reproducing population under the basic forces of mutation and
selection which works in continuous time is given by the following 
system of non-linear differential equations \cite{CK}
\begin{equation} \label{paramuse}
\dot{p}_{\bm{\sigma}}^{}(t) = 
\big( r_{\bm{\sigma}}^{} - \bar{r}(t)\big) p_{\bm{\sigma}}^{}(t)
+ \sum_{\bm{\sigma'}} m_{\bm{\sigma}\bm{\sigma'}} p_{\bm{\sigma'}}(t)\;.
\end{equation}
Here, $p_{\bm{\sigma}}^{}(t)$ denotes the relative frequency of genotype 
${\bm \sigma}$ at time $t$ with corresponding Malthusian fitness
(replication rate minus death rate) $r_{\bm \sigma}^{}$, and
\begin{equation}
\bar{r}(t) = \sum_{\bm{\sigma}} r_{\bm{\sigma}} p_{\bm{\sigma}}(t)
\end{equation}
is the {\em mean fitness} of the population. It is the origin of the
non-linearity in (\ref{paramuse}). Finally, 
$m_{{\bm \sigma}{\bm \sigma'}}$ is the (time independent) rate at which 
${\bm \sigma'}$ mutates to ${\bm \sigma}$. This framework has
originally been defined in classical population genetics \cite{CK}. In 
the sequence space context, it has been introduced in \cite{B} and has been
called the {\it para-muse} ({\em pa}rallel {\em mu}tation-{\em se}lection) 
model, since it assumes mutation and selection to act independently
and in parallel at each instant of time. 
The model ignores recombination and genetic drift due to finite
population size. Both assumptions can be considered as fairly reasonable
at least in the context of the evolution of viruses or bacteria where 
populations can be huge and recombination is absent, or the
nucleotides are tightly linked. In the following subsections, the
basic processes of mutation and selection shall be described in some detail.

\subsection{Mutation}

We take mutation as a point process acting independently on
all sites, ignoring more complicated mechanisms, such as
insertions or deletions. Molecular mutation rates shall be chosen 
according to the following scheme, known as the {\em Kimura 3 ST
model} in molecular phylogeny \cite{Li,SOWH}: 
\begin{figure}[ht]
\centerline{\epsfysize=27mm \epsfbox{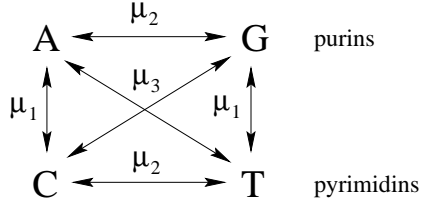}}
\caption{Molecular mutation scheme according to the Kimura 3 ST model.}
\label{mutfig}
\end{figure}

Within this general setup, a number of simpler models is contained,
which treat mutation at different levels of sophistication. In the 
simplest approach, the mutation rates between all four nucleotides 
are assumed to be equal $(\mu_1 = \mu_2 = \mu_3)$. This is the 
so-called {\em Jukes-Cantor mutation scheme}. While this simple 
frame already seems to be sufficient for a number of applications, 
measurements reveal that there are indeed pronounced differences in 
the mutation rates that should be accounted for in more realistic
models. In particular, the {\em transitions} between the two purins 
(A,G) and the two pyrimidins (C,T) are much more frequent than the 
purin--pyrimidin mutations which are called {\em transversions}. This 
may range up to relative differences of 
$\mu_1 \approx \mu_3 \simeq \mu_2/2$ in the 
nucleus and $\mu_1 \approx \mu_3 \simeq \mu_2/40$ in mitochondrial
DNA \cite{Li}. A mutation scheme with $\mu_2 > \mu_1 = \mu_3$ is known as the 
{\em Kimura 2 parameter model}. The full {\em Kimura 3 ST} scheme,
finally, also accounts for the small difference between $\mu_1$ and
$\mu_3$, such that $\mu_2 > \mu_1 > \mu_3$.

Implementing this mutation model into the evolution equation
(\ref{paramuse}), we obtain the following mutation rates between
genotypes ($i \in \{1,2,3\}$)
\begin{equation} \label{mss}
m_{{\bm \sigma}{\bm \sigma'}} = \left\{ 
\begin{array}{rl} 
\mu_i, \quad & d_i({\bm \sigma},{\bm \sigma'})
= d_{{\bm \sigma}{\bm \sigma'}} = 1
\\
-N \sum_i \mu_i,\quad   & {\bm \sigma} = {\bm \sigma'}
\\
0,\quad       & d_{{\bm \sigma}{\bm \sigma'}} > 1
\end{array} \right. \;.
\end{equation}
Here,
\begin{eqnarray} \nonumber
d_1({\bm \sigma},{\bm \sigma'}) & = & 
\#_{A \rightleftarrows C}({\bm \sigma},{\bm \sigma'}) 
+ \#_{G \rightleftarrows T}({\bm \sigma},{\bm \sigma'})
\\ \label{Hamming}
d_2({\bm \sigma},{\bm \sigma'}) & = & 
\#_{A \rightleftarrows G} ({\bm \sigma},{\bm \sigma'})
+ \#_{C \rightleftarrows T}({\bm \sigma},{\bm \sigma'})
\\ \nonumber
d_3({\bm \sigma},{\bm \sigma'}) & = & 
\# _{A \rightleftarrows T}({\bm \sigma},{\bm \sigma'}) 
+ \#_{C \rightleftarrows G}({\bm \sigma},{\bm \sigma'})
\end{eqnarray}
are restricted Hamming distances between ${\bm \sigma}$ and ${\bm \sigma'}$.
In (\ref{Hamming}), $\#_{X \rightleftarrows Y}({\bm \sigma},{\bm \sigma'})$ 
counts the positions at which $X$ and $Y$ are exchanged in $\bm{\sigma}$ and
$\bm{\sigma}'$. Finally,
\begin{equation}
d_{{\bm \sigma}{\bm \sigma'}} = d_1({\bm \sigma},{\bm \sigma'})
 + d_2({\bm \sigma},{\bm \sigma'}) + d_3({\bm \sigma},{\bm \sigma'})
\end{equation}
is the total Hamming 
distance. Note that the choice of the diagonal term 
$m_{{\bm \sigma}{\bm \sigma}}$ in (\ref{mss}) just accounts for 
probability conservation ($\sum_{\bm{\sigma}} 
\dot{p}_{\bm{\sigma}} = 0$) in the mutation part of the 
evolution equation (\ref{paramuse}).

\subsection{Selection and fitness landscape} 

Whereas the mutational part of the dynamics is fairly well understood 
at least on the microscopic (molecular) level, the relation of 
genotype and fitness, which defines the respective selective success, 
is notoriously complex.  
Following the standard notion in molecular evolution, we define the 
{\em fitness function} (or {\em fitness landscape})
\begin{equation}
f: \bm{\sigma} \mapsto r_{\bm{\sigma}} 
\end{equation}
as a mapping from the configuration space $V= \{A,C,G,T\}^N$ into the 
real numbers, assigning a reproduction rate (Malthusian fitness value) 
$r_{\bm{\sigma}}$ to each 
genotype. Implicitly, the fitness function incorporates all the
complicated interactions between the sites. These interactions
are typically long-ranged (since RNA strands or proteins fold in three 
dimensions), highly correlated, and give rise to rather rugged landscapes. 
Especially in the context of RNA evolution, the construction and
characterization of fitness landscapes has motivated numerous studies,
see e.g.\ \cite{Sta} for a review. 

\begin{figure}[t]
\centerline{\epsfysize=50mm \epsfbox{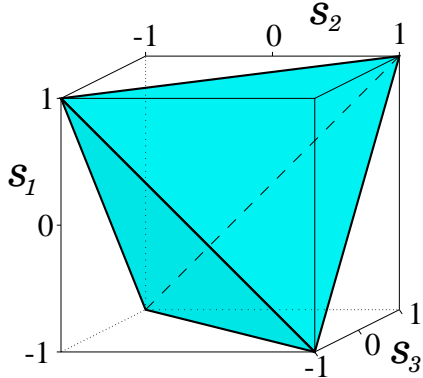}}
\caption{Permutation invariant configuration space of the four-state
  model in surplus coordinates.}
\label{select}
\end{figure}

Below we will show how the evolution equation (\ref{paramuse}), with
an arbitrary choice of the fitness function, can be adapted to the
methods from statistical physics by a reformulation in a quantum 
chain framework. As an application, we then present a study (including
analytical and numerical results) for specific examples from the class
of permutation invariant fitness functions. Here, due to equivalence of
all sites, the fitness of a given genotype is solely a function of 
its restricted Hamming distances from the so called {\em wildtype} sequence
with optimal fitness which we choose as the reference genotype. 
This particularly simple class of fitness 
landscapes is widely used, as a canonical first approximation, 
especially in {\em multilocus theory}. Also in the context of sequence 
space evolution, fitness functions of this type
have been used in a number of studies on the two-state model 
\cite{OB,Leut2,Tara,BBW,WBG}. To implement the approach in our
four-state model, we fix an arbitrary sequence, denoted by 
$\bm{\sigma}_{++}$, as
the wildtype. We will only consider directional selection here towards a
unique genotype with optimal fitness. The fitness of any other
sequence is then determined by the restricted Hamming distances 
$d_i$ relative to $\bm{\sigma}_{++}$. 
Permutation invariance with respect to the position in the sequence
thus leads to a drastic reduction of dimensions. For the four-state
model, the effective configuration
space forms a tetrahedron in 3d (see Fig.~\ref{select}) and is 
conveniently represented in Cartesian coordinates which we 
shall call (following \cite{BBW}) the {\em surplus components}:
\begin{eqnarray}\nonumber
s_1(\bm{\sigma}) &=& 1 - \frac{2}{N}
\Big(d_1(\bm{\sigma},\bm{\sigma}_{++})+d_3(\bm{\sigma},\bm{\sigma}_{++})\Big)\;;
\\ \label{surplus}
s_2(\bm{\sigma}) &=& 1 - \frac{2}{N}
\Big(d_2(\bm{\sigma},\bm{\sigma}_{++})+d_3(\bm{\sigma},\bm{\sigma}_{++})\Big)\;;
\\ \nonumber
s_3(\bm{\sigma}) &=& 1 - \frac{2}{N}
\Big(d_1(\bm{\sigma},\bm{\sigma}_{++})+d_2(\bm{\sigma},\bm{\sigma}_{++})\Big)\;.
\end{eqnarray} 
With this choice, any unstructured random sequence has coordinates 
$s_i \equiv 0$ (with probability 1 in the limit $N\to \infty$).
Any positive value of a surplus component, on the other hand, signals a 
non-trivial overlap of the sequence with the wildtype $\bm{\sigma}_{++}$. 
In particular, $s_1$ measures the surplus of sites with purins or pyrimidins 
as given in $\bm{\sigma}_{++}$ over the purin--pyrimidin mutated sites.

Within this frame, a natural class of permutation invariant fitness 
functions is
\begin{equation} \label{fit}
f: \bm{\sigma} \mapsto
r_{\bm{\sigma}} = N \sum_{i=1}^3 \left[\alpha_i^{} s_i(\bm{\sigma}) + 
\frac{\gamma_i^{}}{2} s_i^2(\bm{\sigma}) \right]
\end{equation}
which includes the following special cases
\begin{itemize}
\item 
Setting $\alpha_i > 0$ and $\gamma_{i} = 0$, we obtain the purely additive 
{\em Fujiyama landscape} without genetic interactions. Here, every 
mutation relative to the wildtype has a fixed deleterious effect, 
independent of any other mutation that may be present in the sequence. 
The additive landscape is a canonical zeroth-order approximation, ignoring
any kind of genetic interactions. In the context of sequence
evolution, this fitness function has been discussed e.g.~in \cite{OB,BBW}. 
\item 
With the choice $\alpha_i \ge - \gamma_{i} > 0$, the model 
corresponds to a concave quadratic fitness function 
(with directional selection) as it is frequently met
in multilocus theory. Due to the gene interactions, existing mutations 
tend to aggravate further ones, which is called {\em positive epistasis}. 
\item
For $\alpha_i \ge 0$ and $\gamma_i > 0$, we finally obtain a convex fitness 
function for directional selection with long-range gene interactions and
{\em negative epistasis} (existing mutations tend to alleviate further 
ones). Since we want to have $\bm{\sigma}_{++}$ as unique wildtype
sequence and a fitness function which is monotonous in the surplus
components, we restrict $f$ to the octant $s_i \ge 0$ and (smoothly) 
truncate the fitness function by introduction of a step function 
$\Theta(s_i)$ whenever frequencies of genotypes with $s_i < 0$ are 
non-zero:
\begin{equation} \label{fit2}
\tilde{f}: \bm{\sigma} \mapsto
r_{\bm{\sigma}} = N \sum_{i=1}^3 
\left[\left(\alpha_i^{} s_i(\bm{\sigma}) + 
\frac{\gamma_i^{}}{2} s_i^2(\bm{\sigma}) \right)\Theta(s_i) \right]\;.
\end{equation}
\end{itemize}
The variables $\alpha_i$ and $\gamma_i$ may further be used to 
distinguish between the effects of the different types of mutations 
(as defined in Fig \ref{mutfig}) on the fitness. In this article, 
we will present explicit results for the two following cases:
\begin{enumerate}
\item 
For the simplest choice, $\alpha_1=\alpha_2=\alpha_3$ and 
$\gamma_1=\gamma_2=\gamma_3$, any mutation away from the wildtype has 
the same effect. Together with the Jukes-Cantor mutation scheme,
symmetry here leads to equal values of the surplus components in the 
mutation--selection equilibrium. The model may thus also be thought
of as a two-state model, where any site is only regarded as occupied 
either with a {\em wildtype} or with a {\em mutant} nucleotide. 
In contrast to the simple two-state model of \cite{BBW}, however, 
there is an effectively asymmetric mutation rate between wildtype 
and mutant in the case considered here.
\item
In a more refined model, we distinguish between transitions and
transversions. In the mutational part, this is done by applying the 
Kimura 2 parameter mutation scheme. In the fitness function, we take 
into account that the deleterious effects of the transversions often 
dominate over those of the transitions: $\alpha_1 > \alpha_{2,3}$ 
and/or $\gamma_1 > \gamma_{2,3}$. 
\end{enumerate}

\section{Quantum chain model}

\subsection{Symmetries}

Since mutation is a random process that is independent of 
the fitness values of the genotypes involved, the molecular mutation
scheme consequently makes no reference to fitness concepts like the
{\em wildtype}. Biological observables measurable from sequence data,
such as the surplus components (\ref{surplus}), and also the fitness 
functions as defined in (\ref{fit}) or (\ref{fit2}), on the other
hand, are defined relative to the wildtype sequence. In order to set 
up these concepts in a common framework, it is convenient to
reformulate also the mutational part of the evolution equation in
coordinates relative to the wildtype. This may always be done  
due to certain symmetries inherent in the mutation scheme of 
Fig.~\ref{mutfig}.

The basic symmetry of the mutation scheme, if all three mutation rates 
$\mu_1, \mu_2, \mu_3$ are pairwise different, is $C_2 \times C_2$ 
(Klein's 4-group), generated by two involutions. If we write the 
operations in standard permutation notation, we can take as generators 
the transformations
\begin{equation}
\begin{pmatrix}
A&C&G&T \\ C&A&T&G
\end{pmatrix} \quad \text{and} \quad
\begin{pmatrix}
A&C&G&T \\ G&T&A&C
\end{pmatrix}\;,
\end{equation}
both being the product of two transpositions. This symmetry may 
now be exploited for a redefinition of the mutation scheme in 
wildtype coordinates. To this end, we fix, for every site of the
wildtype sequence, the element of the 4-group (in the above 
representation) with the letter of the wildtype nucleotide in the
first position (e.g. the string $(T,G,C,A)$ for wildtype nuceotide
$T$). An alternative representation of the configuration space in wildtype
coordinates as
\begin{equation}
{\bm \sigma} \in V^\pm \equiv V_1^\pm \times V_2^\pm
 \times \dots \times V_N^\pm \;;\quad
V_i^\pm = \{++,-+,+-,--\}
\end{equation}
is now given by the mapping, on each site, of the string of 
labels $(++,-+,+-,--)$ to the symmetry element of 4-group defined 
above. With this notation, the three types of mutations included in the 
Kimura 3 ST scheme simply switch the signs of the labels: 
$\pm\pm \to \mp\pm$ at rate $\mu_1$, $\pm\pm \to \pm\mp$ at rate 
$\mu_2$, and $\pm\pm \to \mp\mp$ at rate $\mu_3$.

Higher symmetries of the mutation model are obtained if mutation rates are
equal. For the Kimura 2 parameter scheme, $\mu_1 = \mu_3 \neq \mu_2$,
the operation 
\begin{equation}
A \to C \to G \to T \to A \; = \; 
\begin{pmatrix}
A&C&G&T \\ C&G&T&A
\end{pmatrix}
\end{equation}
is also a symmetry and generates a cyclic group $C_4$. Together with
the previous $C_2 \times C_2$, this generates a dihedral group, $D_4$,
with 8 elements. Finally, if $\mu_1 = \mu_2 = \mu_3$, we additionally
get the simple transposition $A \leftrightarrow C$
and have the full permutation group $S_4$ as symmetry. Note that
$S_4$, which corresponds to the full tetrahedral group with 24
elements, is also the symmetry group of the configuration space of
permutation invariant configurations visualized in
Fig.~\ref{select}. The {\em global} symmetry (with the same
transformation acting at each site simultaneously) of our class of
mutation-selection models with fitness functions according to
(\ref{fit}) is therefore always a subgroup of $S_4$. 
In particular, the symmetric fitness model with $\alpha_1 = \alpha_2 =
\alpha_3$, $\gamma_1 = \gamma_2 = \gamma_3$, and Jukes-Cantor mutation
scheme possesses $C_{3v}$ symmetry, or the full tetrahedral symmetry if the
linear part in the fitness function vanishes ($\alpha_i = 0$).
The transition-transversion model finally, with $\alpha_1 >
\alpha_2 = \alpha_3$, or $\gamma_1 > \gamma_2 = \gamma_3$, and Kimura 2
parameter mutation has simple $C_2$ symmetry, or $D_4$ symmetry if
$\alpha_i \equiv 0$. In the latter case, the combination of
$\gamma_2=\gamma_3$ with $\mu_1=\mu_3$ is necessary, not a
misprint. Other combinations with global $D_4$ symmetry are $(\gamma_1
= \gamma_3; \mu_2=\mu_3)$ and $(\gamma_1=\gamma_2; \mu_1=\mu_2)$. 

\subsection{Construction}

With the above preparations, we may now follow the lines of
\cite{BBW,WBG} where the two-state model is treated.

In a first step, we represent the $4^N$-dimensional vector space in
which we describe the
genotype frequencies as the $N$-fold tensor product space
$W = \otimes_{j=1}^N W_j$. Hereby, the configuration space $V^\pm$ is 
canonically embedded in $W$ by the mapping of the elements of 
$V_i^\pm$ onto the basis vectors 
$\{e_{j}^{++}, e_{j}^{-+}, e_{j}^{+-}, e_{j}^{--}\}$ of $W_j \simeq \RR^4$.
Since the nonlinear part in the differential 
equations (\ref{paramuse}) only amounts to normalization of the 
frequencies, a transformation to so-called
{\em absolute frequencies} \cite{TM,BBW}
\begin{equation}
z_{\bm \sigma}^{}(t) = p_{\bm \sigma}^{}(t) \exp\Big( \sum_{\bm \sigma'} 
r_{\bm \sigma'}^{} \int_0^t p_{\bm \sigma'}^{}(\tau) \,d\tau \Big)  
\end{equation}
then reduces the system to the linear equation
\begin{equation} \label{LGS}
\dot{z}_{\bm \sigma}^{}(t) = \big({\cal M} + {\cal R}\big) 
z_{\bm \sigma}^{}(t) 
\end{equation}
where the mutation and reproduction matrices, ${\cal M} = 
(m_{\bm\sigma \bm\sigma'})$ and ${\cal R} = \text{diag}(r_{\bm\sigma}^{})$, 
may now be conveniently represented in the frequency space $W$. Defining
\begin{equation}
\sigma_j^{(\alpha,\beta)} := \left(\otimes^{j-1} \Id_4 \right) \otimes 
\left(\sigma^\alpha \otimes \sigma^\beta \right)
\otimes \left(\otimes^{N-j-1} \Id_4\right)
\end{equation}
where $\sigma^\alpha$, $\alpha \in \{0,x,z\}$, are the real Pauli matrices and 
$\sigma^0 \equiv \Id_2$, we find
\begin{equation}
{\cal M} = \sum_{j=1}^N \left[ \mu_1 \sigma_j^{(x,0)} + \mu_2 
\sigma_j^{(0,x)} + \mu_3 \sigma_j^{(x,x)} - (\mu_1+\mu_2+\mu_3) \Id\right]
\end{equation}
for the mutation matrix. The reproduction matrix ${\cal R}$ is, for a 
general fitness landscape, an element of the algebra generated by
$\sigma_j^{(z,0)}$ and $\sigma_j^{(0,z)}$, $1\le j\le N$, 
\begin{equation}
{\cal R} = r_0 \Id + \sum_{k,\ell = 1}^N
\sum_{[j_1^{} \dots j_k^{}]} \sum_{[j_1^{} \dots j_\ell^{}]} 
\varepsilon_{[j_1^{} \dots j_k^{}],[j_1^{} \dots j_\ell^{}]}^{}
\prod_{m=1}^k \sigma_{j_m^{}}^{(z,0)} \prod_{n=1}^\ell
\sigma_{j_n^{}}^{(0,z)},
\end{equation}
where $[j_1^{} \dots j_k^{}]$ is an ordered $k$-tupel in $\{1,\dots,N\}$.
Now, from a physical point of view, ${\cal H} = {\cal M} + {\cal R}$
is (up to a global minus sign) the Hamiltonian of two coupled Ising 
quantum chains in a tunable transverse magnetic field (the mutation)
and general spin-interactions within the chains. 

Translated to our quantum chain model, the fitness function of the
permutation invariant landscape defined in (\ref{fit}) results in a 
(longitudinal) magnetic field and a mean field spin-interaction. We find 
${\cal R } = {\cal R}_\alpha + {\cal R}_\gamma$, where
\begin{equation}
{\cal R}_\alpha = \sum_{j=1}^N \left[\alpha_1 \sigma_j^{(z,0)} 
+ \alpha_2 \sigma_j^{(0,z)} + \alpha_3 \sigma_j^{(z,z)} \right]
\end{equation}
and
\begin{equation} \label{rgamma}
{\cal R}_\gamma = \frac{1}{2N} \sum_{j,k = 1}^N \left[ \gamma_1
\sigma_j^{(z,0)}\sigma_k^{(z,0)} + \gamma_2 \sigma_j^{(0,z)}\sigma_k^{(0,z)} +
\gamma_3 \sigma_j^{(z,z)}\sigma_k^{(z,z)} \right]
\end{equation}
Let us stress that, in contrast to most physical applications, the mean 
field model is a much more natural approach in the biological 
context where interactions are typically long-range. So, it is a
legitimate model here, not an inevitable approximation.

\subsection{Biological and physical observables} \label{bpo}

In this subsection, we relate the quantities of biological interest,
mean and variance of the surplus components and the fitness, to the 
physical observables. In what follows, we assume the occuring limits
to exist. 

\paragraph{Genotype composition}
According to (\ref{LGS}), the Hamiltonian of the quantum chain determines the
time evolution of our population of genotypes in an environment that does not 
constrain the population size. For any genotype-independent 
regulation of the population size, the relative genotype frequencies
are found by {\em statistical} normalization. We therefore define the
vector of the genotype composition $|\bm{p}(t) \rangle$ and the 
equilibrium composition $|0\rangle$ as    
\begin{equation}
|\bm{p}(t) \rangle = 
\frac{\exp(t{\cal H})
|\bm{p}_0\rangle} {\langle \Omega|\exp(t{\cal H})|\bm{p}_0\rangle}
\quad ; \quad
|0\rangle := \lim_{t\to \infty} |\bm{p}(t) \rangle 
\end{equation}
where $|\bm{p}_0\rangle$ is the initial composition and
$4^{-N}|\Omega\rangle$ is the equidistribution of genotypes.
Note that the {\em equilibrium composition} of the genotype population
just corresponds to the {\em ground state} of the quantum chain on 
the physical side (with a different `biological' normalization 
$\langle \Omega|0\rangle = 1$).

\paragraph{Fitness} The {\em density of the mean fitness} (or mean
fitness per site) of the population is given by the expression  
\begin{equation}
w(t) := N^{-1} \bar{r}(t) =  
N^{-1} \langle\Omega|{\cal R}|\bm{p}(t)\rangle \;.
\end{equation}
Since
\begin{equation}
w := \lim_{t \to \infty} w(t) = N^{-1} \langle \Omega| {\cal R} | 0
\rangle = N^{-1} \frac{\langle 0| {\cal H} |0\rangle}{\langle 0| 0
  \rangle}
\end{equation}
the {\em equilibrium} mean fitness (per site) is just given by the 
(unique) largest eigenvalue of ${\cal H}$, corresponding to
$|0\rangle$. For an unconstrained population, $w$ also determines the 
growth rate in the long-time limit. In the physical picture, 
$(-w)$ is obviously just the {\em ground state energy} (per spin). 

Using ${\cal M} |\Omega\rangle = 0$, we derive for the time evolution
of the mean fitness
\begin{equation} \label{zeit}
\dot{w}(t) = V_r(t) + N^{-1} 
\langle \Omega| [{\cal R},{\cal M}] | \bm{p}(t) \rangle
\end{equation}
where $V_r(t)$ is the {\em variance of fitness} (per site),
\begin{equation}
V_r(t) = \frac{1}{N}\left(\langle \Omega|{\cal R}^2|\bm{p}(t)\rangle
- \langle \Omega|{\cal R}|\bm{p}(t)\rangle^2 \right)\;.
\end{equation}
In the absence of mutation, (\ref{zeit}) is of course just a special case
of Fisher's ``Fundamental Theorem of Natural Selection'' \cite{Fish} which
states that the rate of increase in fitness is equal to the genetic
variance in fitness. For the mutation-selection models considered
here, the relation has the following intuitive interpretation:
The change in mean fitness is driven by two independent forces. The
first one stems from the change of genotype frequencies due to
selection and is proportional to the variance of fitness values
present in the population. Since variances are positive, it always
tends to increase fitness. The second term on the right hand side of 
(\ref{zeit}) typically decreases fitness. It measures the population
mean of the change in fitness at time $t$ due to the action of mutation.
In mutation-selection equilibrium, both terms balance, and the entire 
residual variance is due to mutation.  

\paragraph{Surplus} Another quantity that characterizes the genetic 
order of the population, as it may be measured from sequence data, is 
the {\em mean surplus}. We define, following and generalizing \cite{BBW},
\begin{equation}
u_i(t) = \sum_{\bm{\sigma}} s_i(\bm{\sigma}) p_{\bm{\sigma}}^{}(t)
\quad ; \quad
u_i = \lim_{t \to \infty} u_i(t) \;. 
\end{equation} 
In particular, 
\begin{equation}
\#_m(t) := \frac{1}{4} \big(3 - (u_1(t)+u_2(t)+u_3(t))\big)
\end{equation}
measures the mean number of mutations per site relative to the wildtype while
\begin{equation}
\#_{tr}(t) := \frac{1}{2} \big( 1 - u_1(t) \big) 
\end{equation}
denotes the mean number of transversions alone.
As a {\em biological order parameter}, the mean surplus plays a 
similar r{\^o}le as the physical magnetization. However, as already 
noted in \cite{BBW2}, both quantities are quite distinct and in many 
cases not even easily related. In the language of the quantum chain, 
the equilibrium mean surplus may be derived as
\begin{equation}
u_1 = \frac{\langle \Omega|\sum_i\sigma_i^{(z,0)}|0\rangle}{N}
\quad ;\quad 
u_2 = \frac{\langle \Omega|\sum_i\sigma_i^{(0,z)}|0\rangle}{N}
\quad ;\quad 
u_3 = \frac{\langle \Omega|\sum_i\sigma_i^{(z,z)}|0\rangle}{N}
\; , 
\end{equation}
whereas the three-component magnetization is defined as the ground 
state expectation value
\begin{equation}
m_1 = \frac{\langle 0|\sum_i\sigma_i^{(z,0)}|0\rangle}
{N \langle 0|0\rangle} \quad;\quad 
m_2 = \frac{\langle 0|\sum_i\sigma_i^{(0,z)}|0\rangle}
{N \langle 0|0\rangle} \quad ;\quad
m_3 = \frac{\langle 0|\sum_i\sigma_i^{(z,z)}|0\rangle}
{N \langle 0|0\rangle} \; .
\end{equation}
As we will show below, magnetization and surplus can show rather
different behaviour especially near phase transitions. The biological 
and physical phase diagrams, however, coincide if phase transitions
(or error thresholds) are defined as nonanalyticity points of the 
ground state energy (or mean fitness) $w$ in the thermodynamic limit
(cf.~the discussion in Section 5). 

\section{Results}

\subsection{Fujiyama model}

As in the two-letter case \cite{BBW}, the quantum chain model 
decomposes into non-interacting one-site Hamiltonians for the 
additive landscape. The mean fitness and its variance are linear
functions in the surplus components. In particular, we obtain from
(\ref{zeit})
\begin{equation}
V_r(t) = \dot{w}(t) + 2\big(
(\mu_1 +\mu_3) \alpha_1 u_1(t)
+ (\mu_2 +\mu_3) \alpha_2 u_2(t) + (\mu_1 +\mu_2) \alpha_3 u_3(t)\big)
\;.
\end{equation}
For Jukes-Cantor mutation, $\mu_1 = \mu_2 = \mu_3 \equiv \mu$, this reduces to
\begin{equation}
V_r(t) = \left(4 \mu + \frac{\text{d}}{\text{d}t}\right) w(t) 
\end{equation}
and $V_r$ is proportional to the mean fitness in the
mutation--selection equilibrium. Exact results are easily
found from the solution of the four-dimensional eigenvalue problem of
the one-site Hamiltonian. We only give the expression for the mean
fitness in the symmetric case, $\alpha_1 = \alpha_2 = \alpha_3 \equiv \alpha$
with Jukes-Cantor mutation scheme ($\mu_1 = \mu_2 = \mu_3 \equiv \mu$):
\begin{equation}
w(t) = 
\frac{\exp[2t(\alpha+\mu)]\cosh[2tQ]\left(\alpha-2\mu+2Q\tanh[2tQ]\right)
-\alpha-4\mu}{1+\exp[2t(\alpha+\mu)]\cosh[2tQ]}
\end{equation}
where
\begin{equation}
Q = \sqrt{\mu^2+\alpha^2 -\alpha\mu} 
\end{equation}
and the equidistribution of genotypes is chosen as starting configuration.

Means and variances of the fitness and the surplus in
mutation--selection balance are shown in Fig.~\ref{finite} below. 
A plot of the time evolution of fitness is given in Fig.~\ref{time2}. 
There is clearly no phase transition (resp.~no {\em error threshold} 
behaviour) for the additive Fujiyama landscape, as expected in view of
the complete absence of interactions (resp.\ epistasis).

%\begin{equation}
%w = \alpha \left(2\sqrt{\left(\frac{\mu}{\alpha}\right)^2 - 
%\frac{\mu}{\alpha} + 1} - 2\frac{\mu}{\alpha} +1\right) 
%\end{equation}

\subsection{Quadratic fitness model: Equilibrium results}

In contrast to the additive case, no simple relation between surplus
and fitness is known in the case of the quadratic landscape as
long as $t$ or $N$ are kept finite. However, due to the permutation
invariance of the Hamiltonian, the individual fitness--surplus
relation (\ref{fit}) is recovered in the thermodynamic limit
for the corresponding mean values of the equilibrium population. 
We obtain in analogy to \cite{BBW2}:
\begin{equation} \label{surrel}
w = \lim_{t \to \infty} w(t) = \sum_{i=1}^3 \left(\alpha_i u_i 
+ \frac{\gamma_i}{2} u_i^2 \right)
\end{equation}
and, from (\ref{zeit}), for the equilibrium variance of fitness per site
\begin{multline} \label{variance}
V_r = \lim_{t \to \infty} V_r(t) = 
2(\mu_1+\mu_3)\left(\alpha_1 u_1 + \gamma_1 u_1^2\right) +
\\
2(\mu_2+\mu_3)\left(\alpha_2 u_2 + \gamma_2 u_2^2\right) + 
2(\mu_1+\mu_2)\left(\alpha_3 u_3 + \gamma_3 u_3^2\right)\;.
\end{multline}
%\begin{equation}
%\textswab{h}= \mu_1\sigma^{(x,0)}+\mu_2 \sigma^{(0,x)} +\mu_3 \sigma^{(x,x)} +%\gamma_1 m_1 \sigma^{(z,0)} + \gamma_2 m_2 \sigma^{(0,z)} + \gamma_3 m_3 
%\sigma^{(z,z)}
%\end{equation}
The key to the solution in the thermodynamic limit is now the minimum
principle of the physical free energy which translates to a maximum
principle for the equilibrium mean fitness. Maximizing
\begin{equation}
\langle \bm{x} | {\cal M} + {\cal R} | \bm{x} \rangle - 
w \big(\langle \bm{x} |\bm{x}  \rangle -1\big)
\end{equation}
with respect to $w$ and $\bm{x}$, we obtain, taking permutation symmetry of 
$\bm{x}$ into account, the following variational expression for $w$:
\begin{equation} \label{fitm}
\begin{align} \nonumber
w(\bm{\alpha},\bm{\mu}&,\bm{\gamma}) \;\; = 
\sup_{m_1,m_2,m_3} \bigg[\alpha_1 m_1 + 
\alpha_2 m_2 + \alpha_3 m_3 + \frac{\gamma_1}{2} m_1^2 +
\frac{\gamma_2}{2} m_2^2 + \frac{\gamma_3}{2} m_3^2 +
\\ \nonumber
&\frac{\mu_1}{2}
\left(\sqrt{(1+m_2)^2-(m_1+m_3)^2}+\sqrt{(1-m_2)^2-(m_1-m_3)^2}-2\right)+
\\ \nonumber
&\frac{\mu_2}{2}
\left(\sqrt{(1+m_1)^2-(m_2+m_3)^2}+\sqrt{(1-m_1)^2-(m_2-m_3)^2}-2\right)+
\\
&\frac{\mu_3}{2}
\left(\sqrt{(1+m_3)^2-(m_1+m_2)^2}+\sqrt{(1-m_3)^2-(m_1-m_2)^2}-2
\right)\bigg]
\end{align}
\end{equation}
where $m_i \in [-1,1]$ are the components of the physical
magnetization. Let us stress that, from the biological point of view, 
the translation to the physical framework seems a necessary technical 
step since we do not know of any variational principle for the 
biological model which works directly in $L^1$. We now take a closer 
look at two special cases.

\paragraph{Symmetric fitness model} For the symmetric 
{\em wildtype--mutant} model with $\alpha_i \equiv \alpha$, 
$\gamma_i \equiv \gamma$ and Jukes-Cantor mutation rate $\mu$, 
all components of the order parameters are equal, 
$m_i \equiv m$ and $u_i \equiv u$, respectively. 
Here, the variational expression (\ref{fitm}) for $w$ leads to the 
following self-consistency condition for $m$:
\begin{equation} \label{sc}
m = \frac{1}{3}\left[ 1 + \frac{2(\alpha + \gamma m) - \mu}
{\sqrt{(\alpha + \gamma m)^2 - \mu(\alpha + \gamma m) + \mu^2}}\right]\;.
\end{equation}
This is a quartic equation in $m$ and can be solved using the
standard formulas. However, since the explicit solution is rather
lengthly, we do not include it here, but give a qualitative
discussion instead.

Obviously, the relation has a unique real solution for any $\alpha$ and
$\mu$ whenever $\gamma$ is {\em negative}. Like in the case of the 
two-state model, we thus obtain no phase transition for positive 
epistasis. In the following, we therefore concentrate our discussion 
on positive $\gamma$ (or negative epistasis). Note that, for
calculations in the thermodynamic limit, always the fitness function $f$
(\ref{fit}), and hence the reproduction matrix ${\cal R_\gamma}$
(\ref{rgamma}), can be used instead of the truncated form $\tilde{f}$
(\ref{fit2}), since the frequencies of genotypes with negative surplus 
vanish. For $\alpha_i \equiv 0$, this is due to spontaneous breaking of 
the extra $C_2 \times C_2$ symmetry of 
${\cal H} = {\cal M} + {\cal R_\gamma}$. 

In contrast to the two-state model, where a phase transition in the
thermodynamic limit is only found for zero external field, it turns
out that the present model has phase transitions for a whole range of
the linear fitness parameter $\alpha$ when epistasis is negative: 
For $\tilde{\alpha} := \alpha/\gamma$ in the interval
\begin{equation}
0 \le \tilde{\alpha} < \frac{1}{3}
\left(\sqrt{\frac{4}{3}}-1\right) \simeq 0.0515668
\end{equation}
we find a first order phase transition of the system at
\begin{equation}
\tilde{\mu} := \frac{\mu}{\gamma} = \tilde{\mu}_c = \frac{2}{3}
 + 2 \tilde{\alpha}
\end{equation}
with a finite jump in the magnetization from $m_+$ to $m_-$ where
\begin{equation}
m_\pm = \frac{1}{3}\left(1 \pm
\sqrt{1 - 27 \tilde{\alpha}^2 - 18\tilde{\alpha}}\right)\;.
\end{equation}
From $m$ we derive the mean fitness $w$ using (\ref{fitm}), from $w$
we obtain the surplus $u$ via (\ref{surrel}) and, finally, the variance of
the fitness $V_r = 12\mu(\alpha u +\gamma u^2)$.
Looking at the surplus $u$, we also find a phase transition at 
$\tilde{\mu}= \tilde{\mu}_c$. As $m$, it vanishes in the disordered
phase for $\alpha = 0$. Note however that, since $w$ is continuous,
due to the relation (\ref{surrel}), also the surplus is continuous at a phase 
transition. In \cite{BBW2} it has been shown that these differences of the
biological and physical order parameters arise with the change from classical
to quantum mechanical probabilities (resp.\ the change from $L^1$ to $L^2$) 
in translating the biological model into the physical one. We remark
that a different, discontinuous behaviour of the biological order 
parameter at a (physical) first order transition has been observed for 
the sharply peaked landscape in Eigen's quasispecies model \cite{FP}.
Mean fitness and its variance, magnetization, and surplus for different
values of $\alpha$ are shown below in Fig.~\ref{JC}.  

\begin{figure}[th]
\centerline{\epsfxsize=65mm \epsfysize=55mm \epsfbox{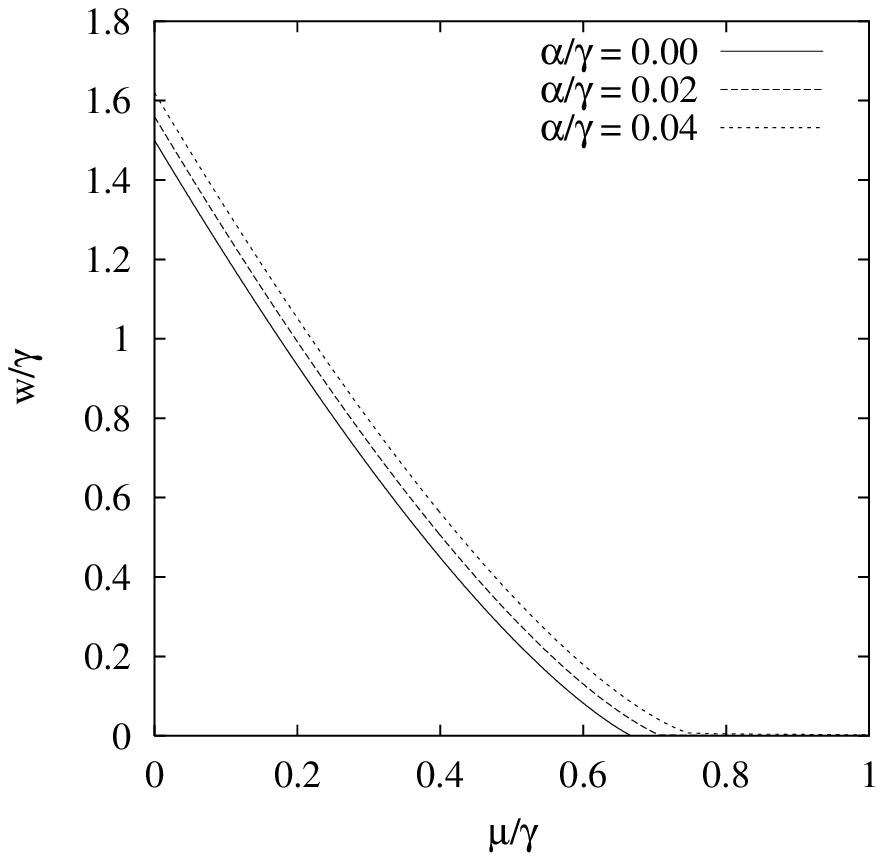} 
\epsfxsize=65mm  \epsfysize=55mm \epsfbox{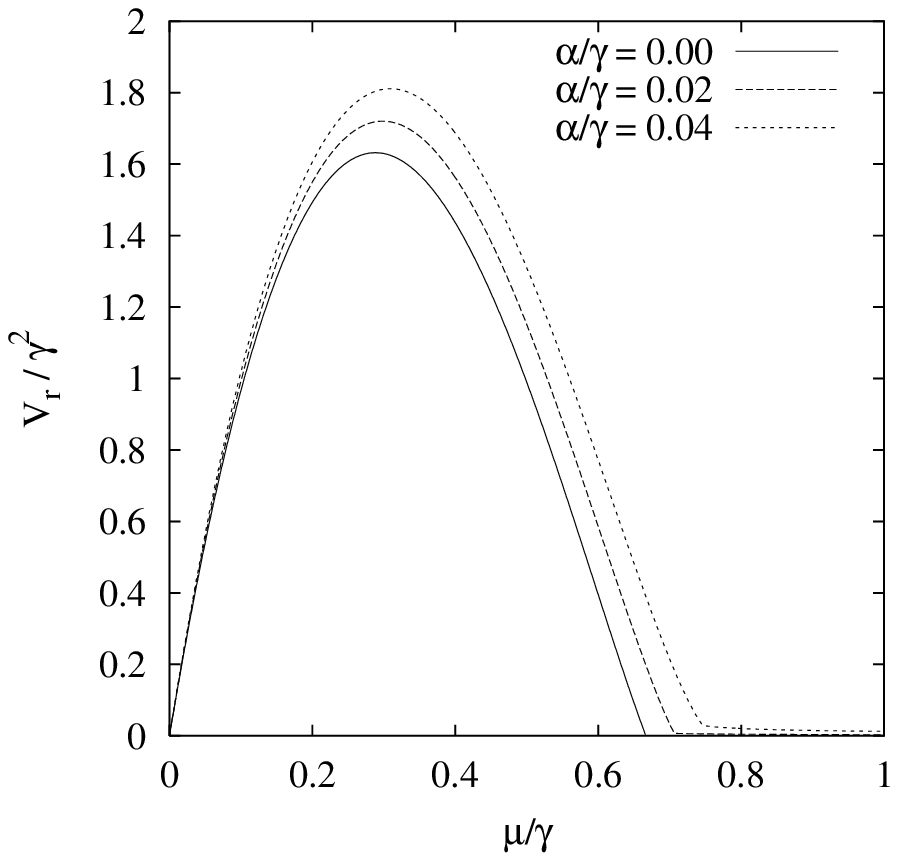}}
\centerline{\epsfxsize=65mm \epsfysize=55mm \epsfbox{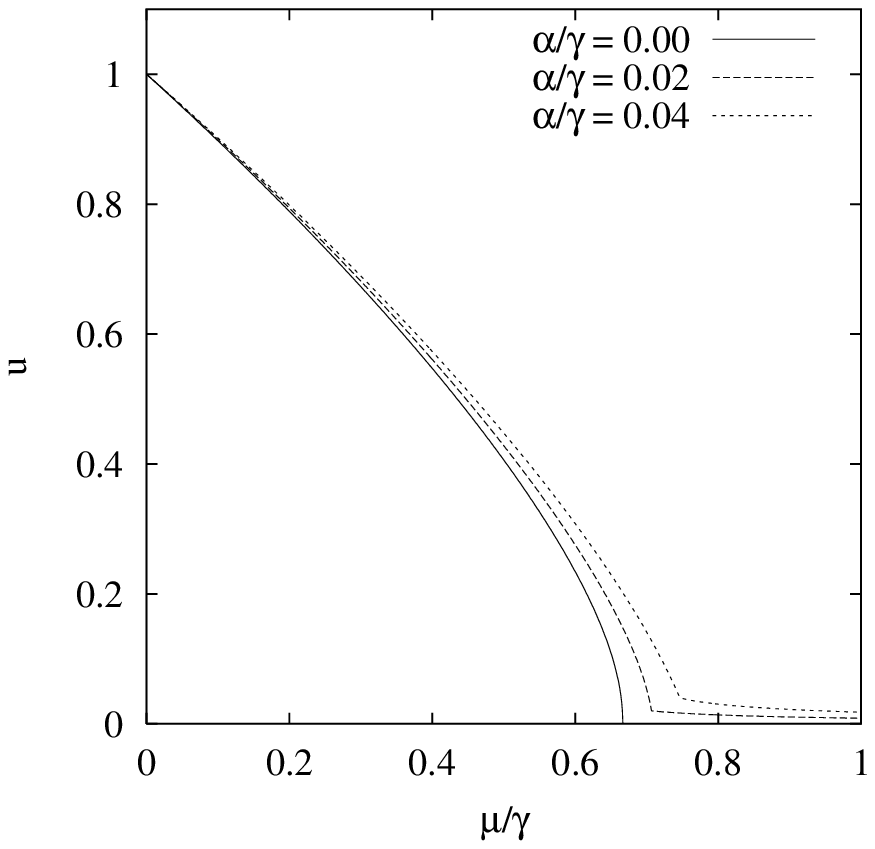} 
\epsfxsize=65mm  \epsfysize=55mm \epsfbox{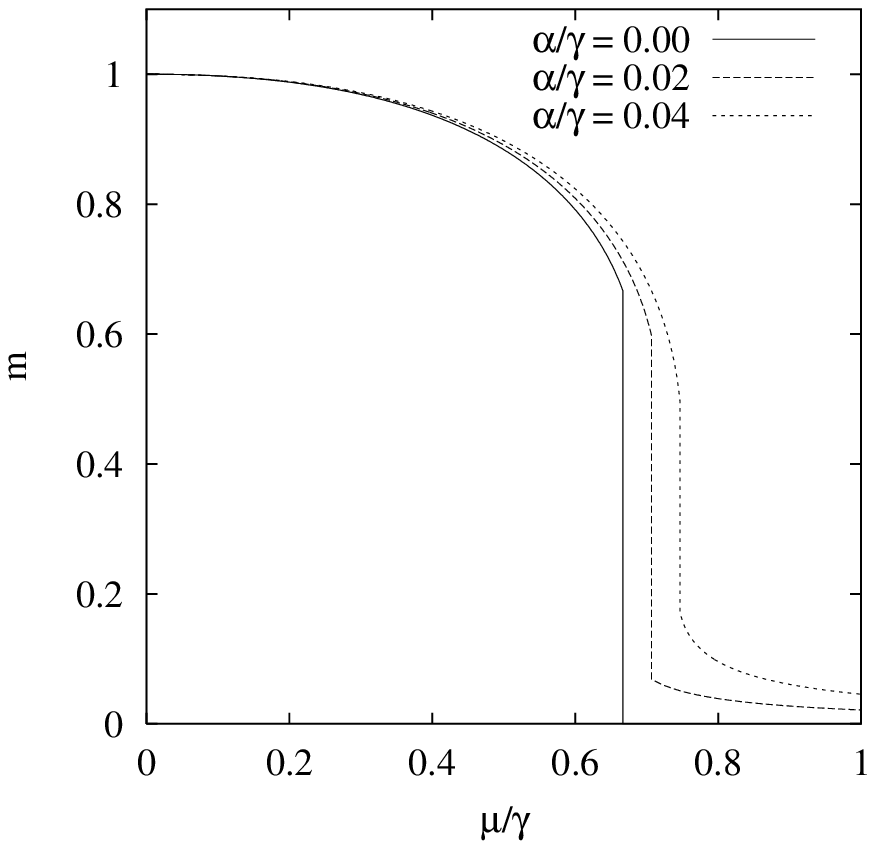}}
\caption{Mean fitness and its variance, surplus and magnetization in
  the symmetric fitness model for various linear parts of the fitness
  function in the infinite sites limit.}
\label{JC}
\end{figure}

\paragraph{Transition--transversion model} In our second example, we
wish to distinguish mutations between like and unlike nucleotides. In 
a first step, we retain the symmetric fitness landscape 
$\gamma_1 = \gamma_2 = \gamma_3 \equiv \gamma$ (for simplicity
with vanishing linear part $\alpha = 0$), but let the relative
frequencies of transitions and transversions differ by assuming the 
{\em Kimura 2 parameter} mutation scheme, 
$\mu_1 = \mu_3 \equiv \mu \neq \mu_2$. 

\begin{figure}[ht]
\centerline{\epsfysize=60mm \epsfbox{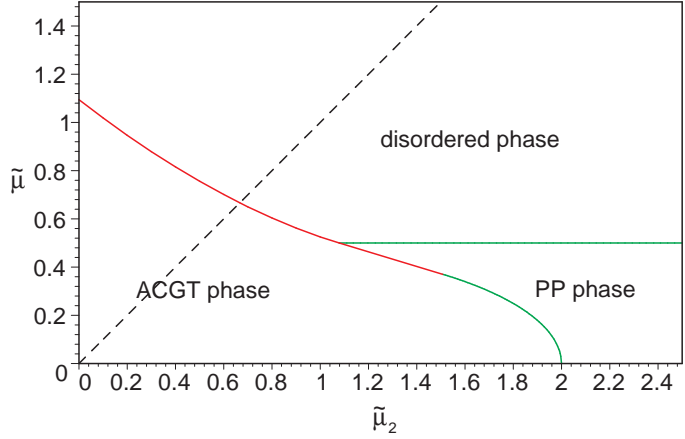}}
\caption{Phase diagram of the transition--transversion model with 
with symmetric fitness landscape and Kimura 2 parameter mutation
scheme. Solid and dotted lines correspond to first and second order
phase transitions, respectively. The dashed line indicates the  
Jukes-Cantor mutation scheme.}
\label{pd1}
\end{figure}
\begin{figure}[ht]
\centerline{\epsfxsize=43mm\epsfysize=35mm \epsfbox{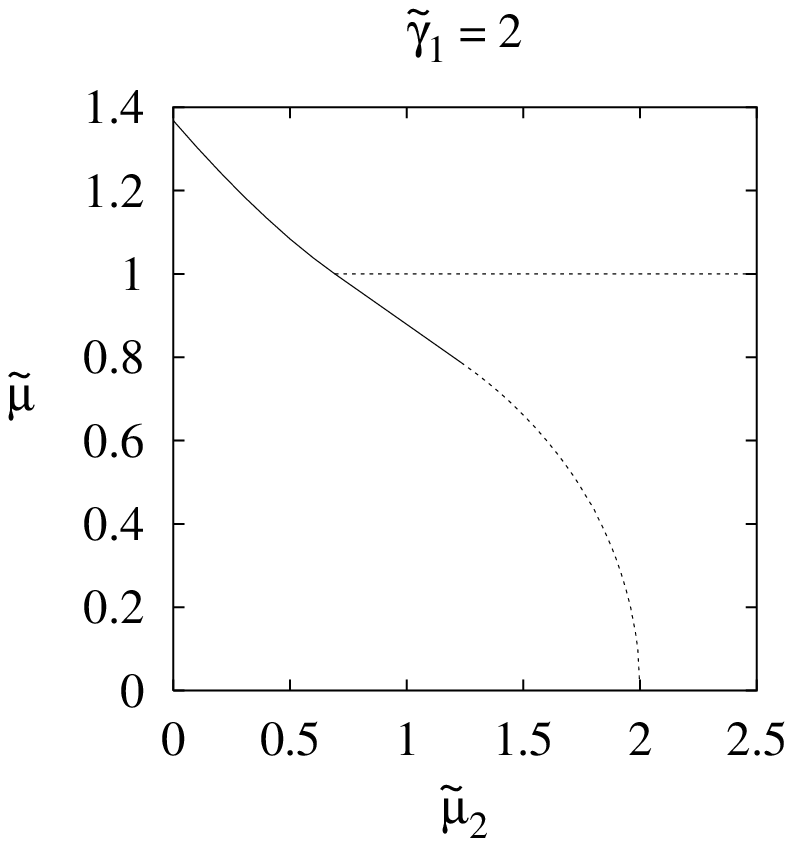}
\epsfxsize=43mm\epsfysize=35mm \epsfbox{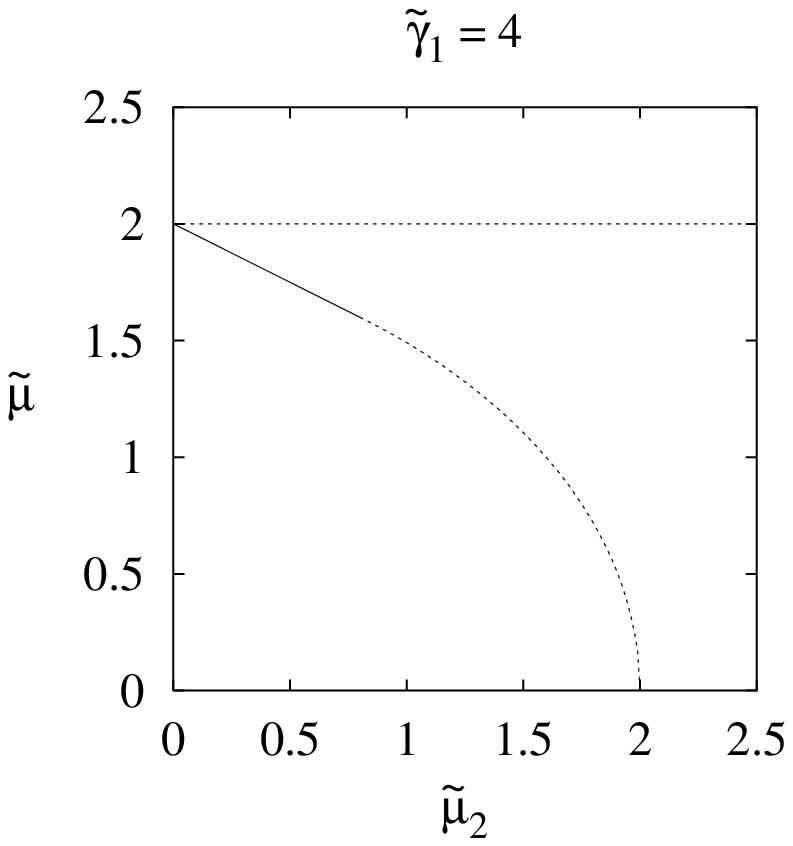}
\epsfxsize=43mm\epsfysize=35mm \epsfbox{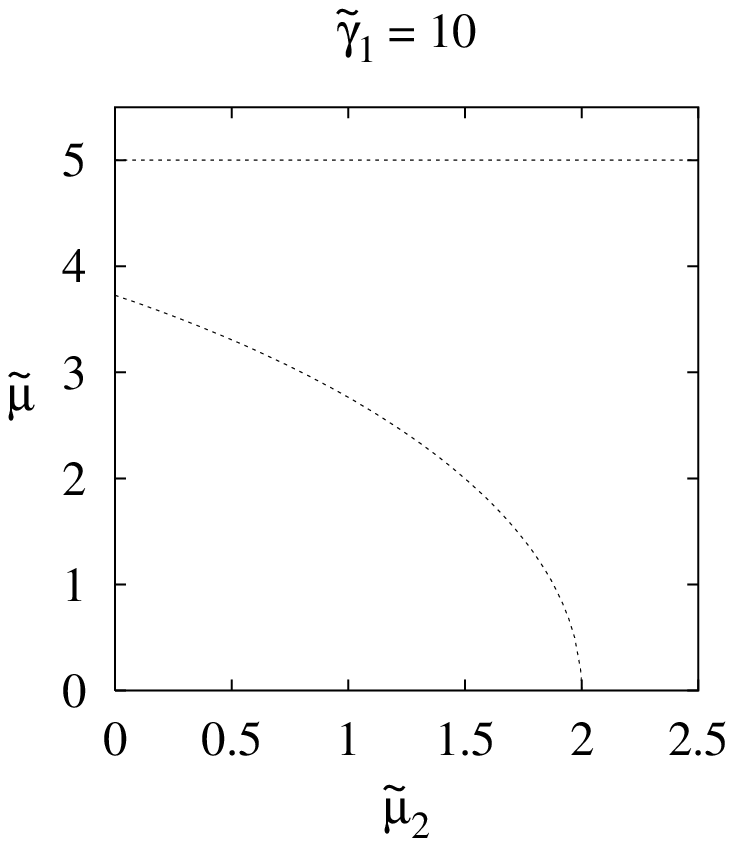}}
\caption{Phase diagrams for anisotropic fitness landscapes $\gamma_1 > 
\gamma_2 = \gamma_3 \equiv \gamma$ and Kimura 2 parameter mutation
scheme. Solid and dotted lines correspond to first and second order
phase transitions, respectively.}
\label{pd2}
\end{figure}

In the extended parameter space of the reduced mutation rates
$\tilde{\mu} = \mu/\gamma$; $\tilde{\mu}_2 = \mu_2/\gamma$, we now 
obtain a phase diagram with {\em three} distinct phases 
(see Fig.~\ref{pd1}).
\begin{itemize}
\item
For $\tilde{\mu}$ and $\tilde{\mu}_2$ sufficiently small, 
all three surplus components 
are positive, indicating genetic order with respect to the entire 
4-letter alphabet of the nucleotides: {\em ACGT phase}.
\item
If we increase the mutation rate $\tilde{\mu}_2$ for low $\tilde{\mu}$, 
the system crosses over to a phase which does no longer distinguish 
between the different kinds of purins (A,G) and pyrimidins (C,T), but 
is still ordered with respect to transversions. This is the limiting
case described by the two-state model. We call this the {\em PP phase}.
\item
For higher mutation rates $\tilde{\mu},\tilde{\mu}_2$, we finally enter a
completely {\em disordered phase} with vanishing fitness and surplus.
\end{itemize}
In a second step, we now also let the mutation effects of transitions
and transversions differ and assume a fitness landscape
with $\gamma_2 = \gamma_3 \equiv \gamma$, but $\gamma_1 \neq \gamma$
in general. The changes in the phase diagram for increasing 
$\tilde{\gamma}_1 = \gamma_1/\gamma$ are shown in Fig.~\ref{pd2}. 
The phase transitions between the three phases may be first or second 
order. In general, we obtain the following phase space structure:
\begin{itemize}
\item
Phase transitions between the disordered and PP phase are second order and
located on the line $\tilde{\mu} = \tilde{\gamma}_1/2$. This phase
transition corresponds to the one also seen in the two-state model \cite{BBW}.
\item
The phase transition line between the ACGT and PP phases in 
general changes from first to second order with increasing 
mutation rate $\tilde\mu_2$ (see Figs.~\ref{pd1}, \ref{pd2}). 
For the second order transitions we derive, on 
expanding (\ref{fitm}) to lowest order in $m_2 = m_3$,
\begin{equation}
\mu = \frac{\gamma_1}{\gamma_1 + 2\gamma}
\sqrt{(\gamma_1 + \mu_2)(2\gamma-\mu_2)} \;.
\end{equation}
Numerically, we find that the first order transitions are 
located on a straight 
line up to $\tilde{\mu} = \tilde{\gamma}_1/2$ where the PP phase
changes into the disordered phase. The $\tilde{\mu}_2$-interval of
first-order transitions decreases for increasing $\tilde{\gamma}_1$. 
For $\tilde{\gamma}_1 \gtrapprox 8.45$, all phase transitions 
between the ACGT and PP phases are second order.
\item
Finally, for $\tilde{\gamma}_1 \le 4$, there are direct first order 
phase transitions between the ACGT phase and the disordered phase 
(for $\tilde{\mu}_2$ sufficiently small). For higher values of 
$\tilde{\gamma}_1$, these two phases are separated by the PP phase.
\end{itemize}

As for the symmetric fitness function discussed above, there are no
compact analytic expressions for the fitness or the surplus in the
ACGT phase. In the PP phase, however, the following values for 
the mean fitness and the non-zero components of the mean surplus and the 
magnetization are found: 
\begin{equation}
w = \frac{\gamma_1}{2} \left(1 - \frac{2\mu}{\gamma_1}\right)^2  \quad ; \quad 
u_1 = 1 - \frac{2\mu}{\gamma_1} \quad ; \quad 
m_1 = \sqrt{1- \left(\frac{2\mu}{\gamma_1}\right)^2}\;. 
\end{equation}
The variance in fitness per site, finally, is proportional to the mean
fitness in the PP phase: $V_r = 8 \mu w$. Note that all these
expressions are independent of the transition rate $\mu_2$ and
directly comparable to the results of the two-state model
\cite{BBW,WBG} by idebtifying $\{++,+-\}$ with `$+$' and $\{-+,--\}$
with `$-$'.

\subsection{Quadratic fitness model: Finite sequence length} \label{fs}

For the Fujiyama model with independent sites, all the quantities
calculated here, means and variances per site in infinite populations, 
are independent of the assumed length $N$ of the sequences.
This is no longer the case for models including epistasis. In this 
subsection, we therefore present a quick numerical investigation of the 
symmetric fitness model
for finite system sizes and compare the results with those in the 
thermodynamic limit. Since the frequencies of genotypes with negative 
values of the surplus no longer vanish for finite sequences, we use
the truncated fitness function (\ref{fit2}), with $\gamma_i \equiv
\gamma > 0$ and $\alpha_i = 0$ for our calculations.

\begin{figure}[ht]
\centerline{\epsfxsize=65mm \epsfysize=55mm \epsfbox{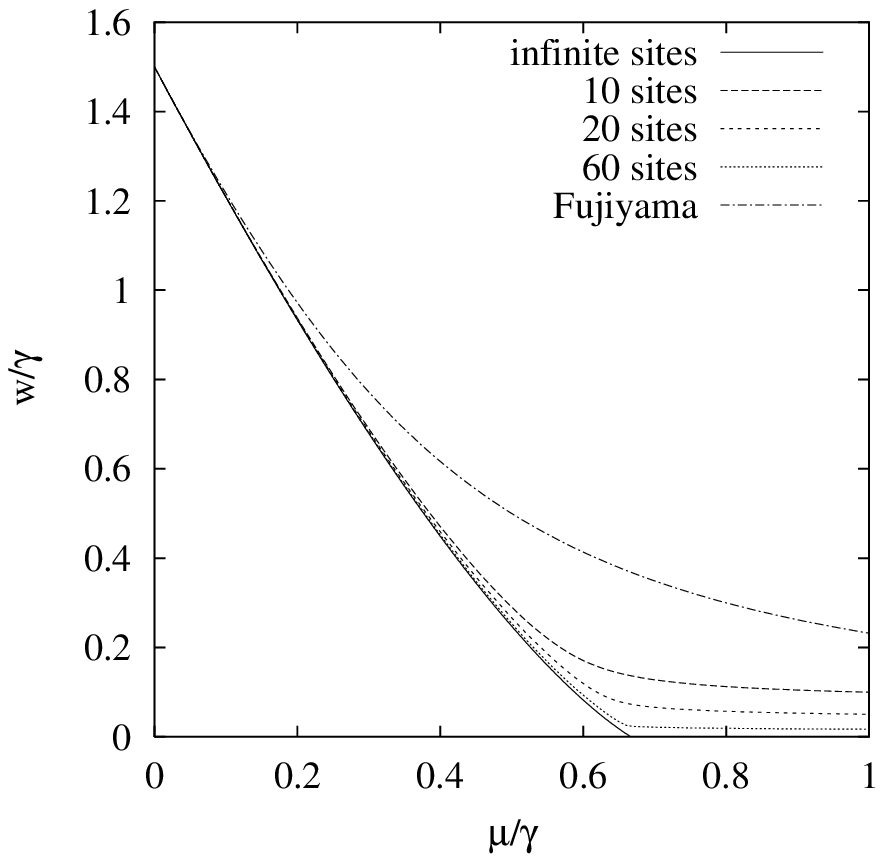} 
\epsfxsize=65mm  \epsfysize=55mm \epsfbox{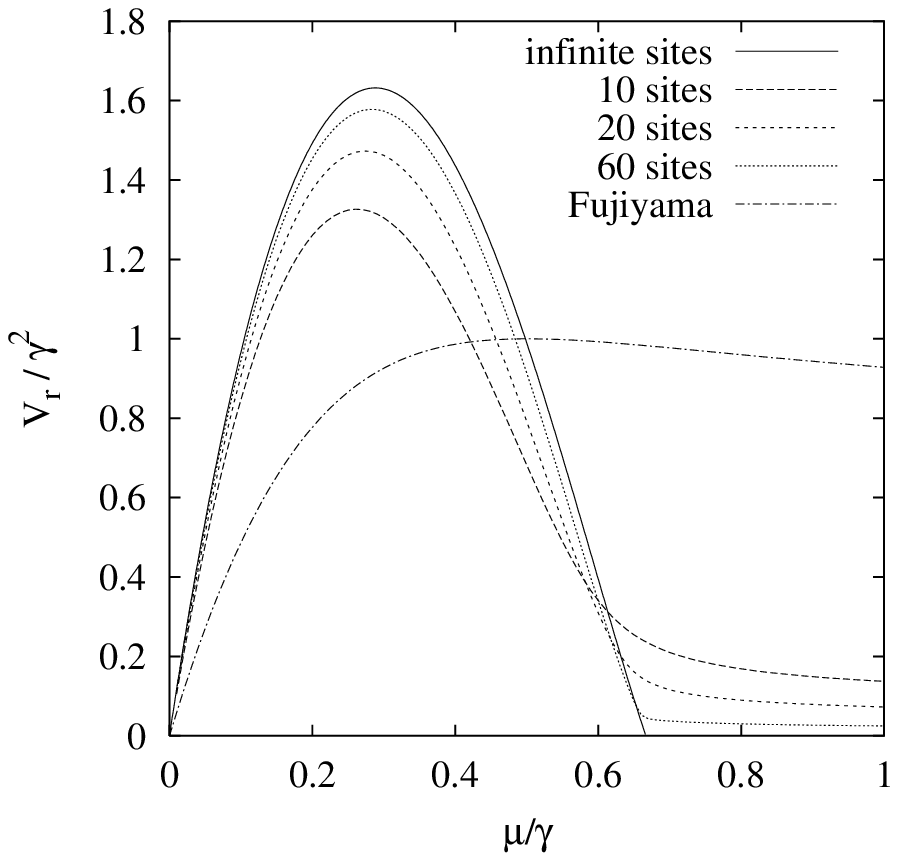}}
\centerline{\epsfxsize=65mm \epsfysize=55mm \epsfbox{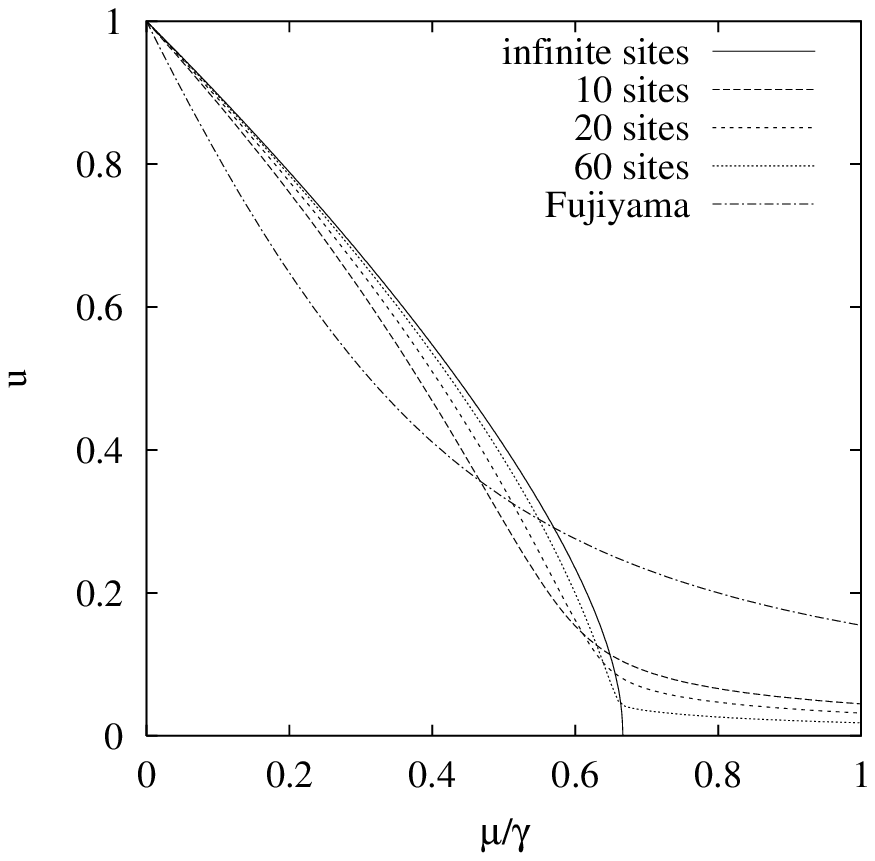} 
\epsfxsize=65mm  \epsfysize=55mm \epsfbox{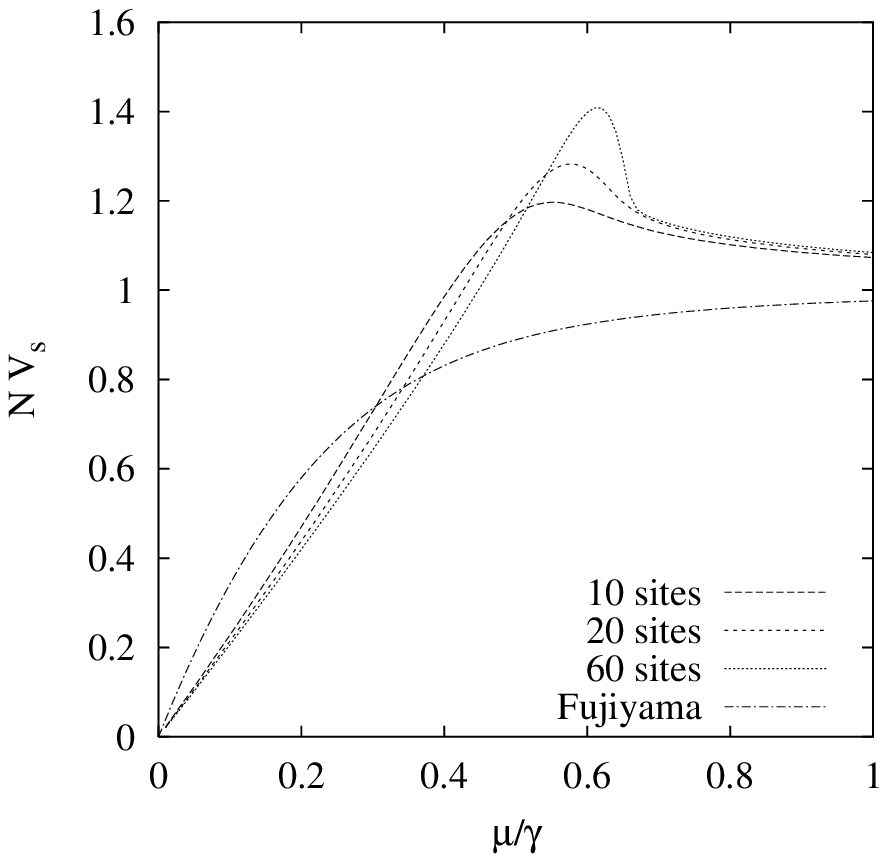}}
\caption{Equilibrium behaviour of fitness and surplus of the symmetric
  fitness model with finite sequence length. Results for the Fujiyama
  model with scaling $\alpha = \gamma/2$ are also shown.}
\label{finite}
\end{figure}

All results are obtained by direct numerical solution of the eigenvalue 
problem in the $[(N+1)(N+2)(N+3)/6]$-dimensional vector space of 
permutation invariant population vectors. Numerically precise
calculations have been performed up to $N = 60$ (39711-dim.), the results
are shown in Fig.~\ref{finite}. It is seen that the mean surplus and
the mean and the variance of the fitness rapidly approach the limiting
curves and behave qualitatively different from the Fujiyama model
even for very small system sizes. We also show the finite-size
behaviour of the variance of the surplus $V_s$. Since this quantity 
vanishes as $1/N$, it is not obtainable from the leading order terms 
in the thermodynamic 
limit. In our finite size calculations, we rescale $V_s$ with the 
sequence length to obtain comparable results. Whereas $V_s$ is
monotonously increasing for the additive model (where $N V_s = 1-
u^2$), it runs through a maximum for quadratic fitness. Note that this 
maximum, in contrast to the variance of fitness, is located directly
at the error threshold. The behaviour is qualitatively similar to the
two-state model \cite{Oli}.

Since there has been some discussion recently on the correct scaling 
of fitness values and mutation rates with the length of the sequence (cf
\cite{FP,BG}), let us finally remark that the finite size results in
this and the next section show that our choice, keeping fitness and
mutation rate {\em per site} fixed, is adequate for all quantities
considered here.

\subsection{Quadratic fitness model: Time evolution}

Originally, the error threshold has been defined as an equilibrium 
phenomenon (cf \cite{ECS,BG}): For special forms of the fitness
landscape, there is a finite critical value $\mu_c$ of the mutation 
rate beyond which genetic order is no longer maintained by selection. 
For the four-state model with quadratic fitness, this situation has been 
discussed above.
However, for a suitable fitness function, the threshold
is not necessarily connected with high mutation rates. 
In this subsection, 
we consider the relaxation of a non-equilibrium population to 
mutation-selection balance. It turns out that, depending on the
starting configuration, an even stronger threshold effect may be 
observed in the time evolution of the fitness and the surplus for
all mutation rates below the critical equilibrium value.

\begin{figure}[ht]
\centerline{\epsfxsize=65mm \epsfysize=55mm \epsfbox{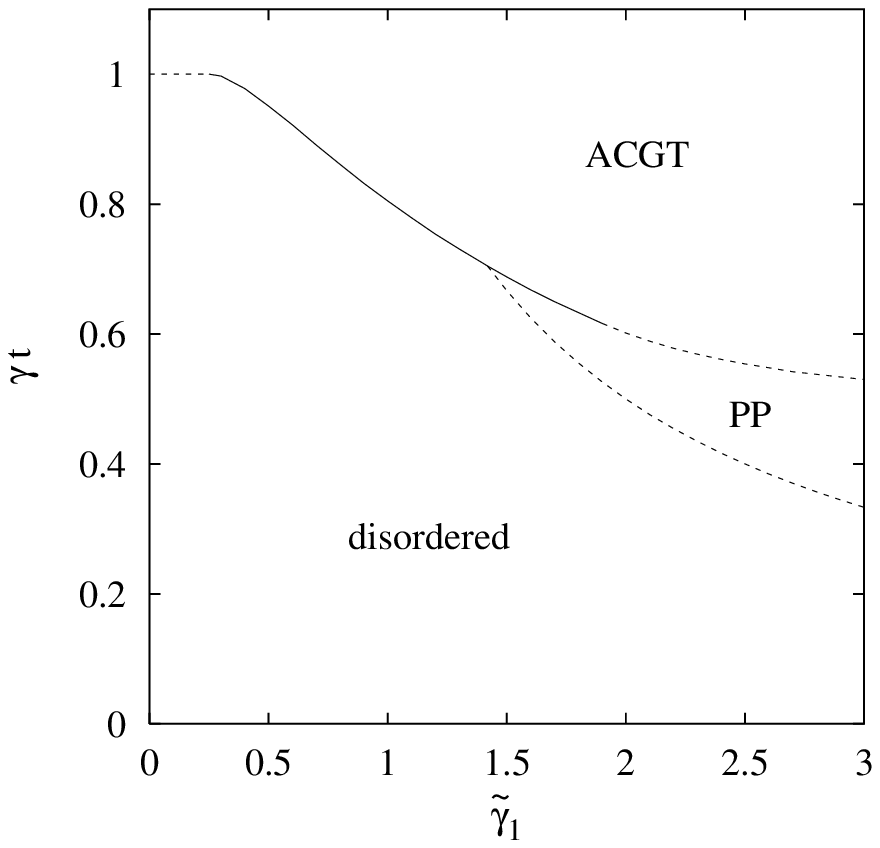}
\epsfxsize=65mm \epsfysize=55mm \epsfbox{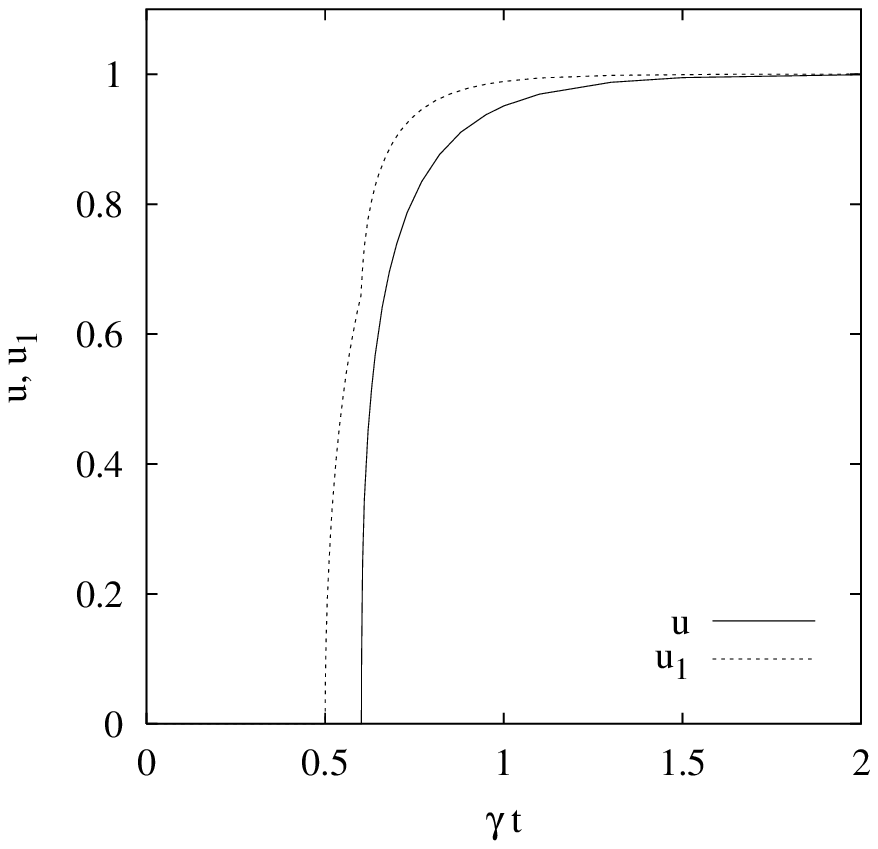}}
\caption{Dynamical phase diagram of the transition-transversion model
  for vanishing mutation starting from the equidistribution. (Solid: 
  first order; dashed: second order transition). Right: Time 
  evolution of the surplus components for $\tilde{\gamma}_1 = 2$.}
\label{time1}
\end{figure}

\paragraph{Zero-mutation limit of the transition-transversion model}
The essence of the threshold phenomenon in the time evolution is
already contained in the selection dynamics alone. In a first step, we
therefore disregard mutation altogether by working in the
zero-mutation limit. Obviously, we then deal with a classical
mean-field model on the physical side. As our starting configuration, 
we choose the completely unstructured population with an equidistribution of
genotypes $|\bm{p}_0\rangle = 4^{-N}|\Omega\rangle$. 
In this particular situation, some progress is possible also
analytically. Noting that
\begin{equation}
\langle \hat{C} \rangle(t) = 
\frac{\langle \Omega|\hat{C} \exp(t {\cal
    R})|\Omega\rangle}{\langle \Omega|\exp(t {\cal R})|\Omega\rangle}
 = \frac{\text{tr}(\hat{C} \exp(t {\cal R}))}{\text{tr}(\exp(t {\cal R}))}
\end{equation}
for any element $\hat{C}$ of the algebra generated by
$\{\sigma_i^{(z,0)},\sigma_i^{(0,z)}\}$, the biological and physical
pictures coincide in this case. Using the fitness function
of the transition-transversion model with 
$\gamma_2 = \gamma_3 \equiv \gamma > 0$, we obtain the
following implicit equations for the time evolution of the surplus 
components:
\begin{eqnarray}
u &=& \frac{\sinh(2\gamma t u)} {\cosh(2\gamma t u) + 
\exp[ -2\gamma_1 t(2u\coth(2\gamma t u) -1)]}
\\[1mm]
u_1 &=& \frac{\cosh[\gamma t Q(u_1)] - \exp(-2\gamma_1 t u_1)} 
{\cosh[\gamma t Q(u_1)] + \exp(-2\gamma_1 t u_1)}
\end{eqnarray}
where
\begin{equation}
Q(u_1) = \sqrt{(1+u_1)^2-\exp(4\gamma_1 t u_1)(1-u_1)^2}\;.
\end{equation}
The resulting dynamical phase diagram is shown in Fig.~\ref{time1}. 
As in the equilibrium situation, there are three phases.
Depending on the ratio $\tilde{\gamma}_1 = \gamma_1/\gamma$, the 
system directly crosses to an ordered phase after a sharply defined 
waiting time $t_c$, or performs two consecutive transitions, entering 
the PP phase in the first one.  

\begin{figure}[ht]
\centerline{\epsfxsize=65mm \epsfysize=55mm \epsfbox{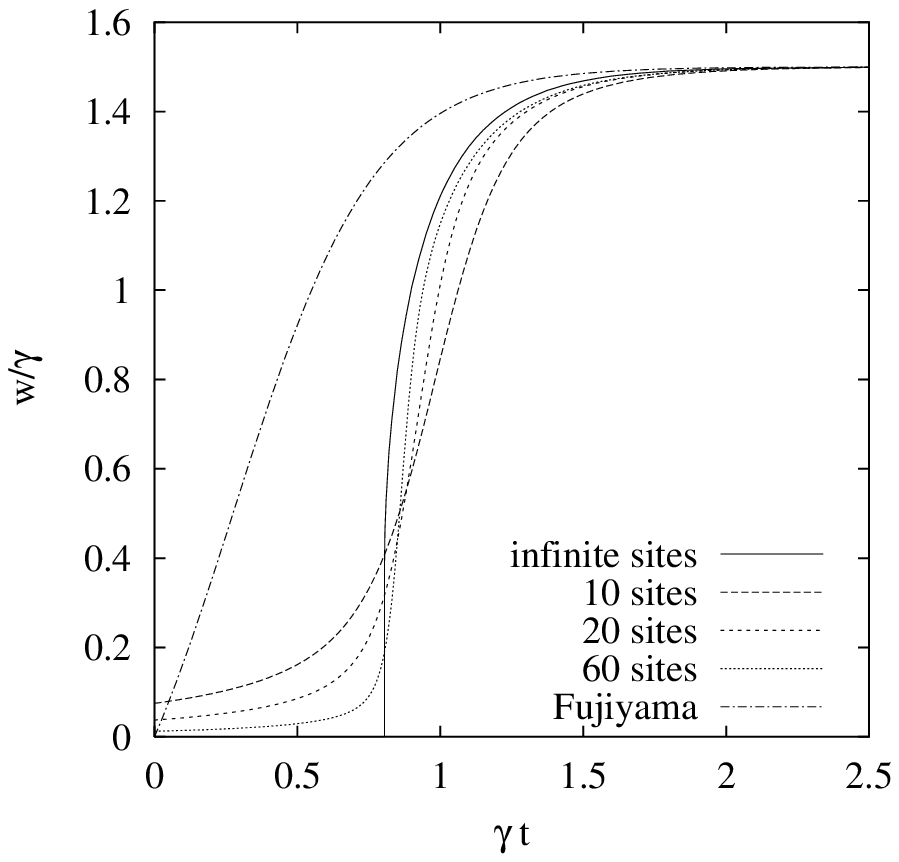} 
\epsfxsize=65mm  \epsfysize=55mm \epsfbox{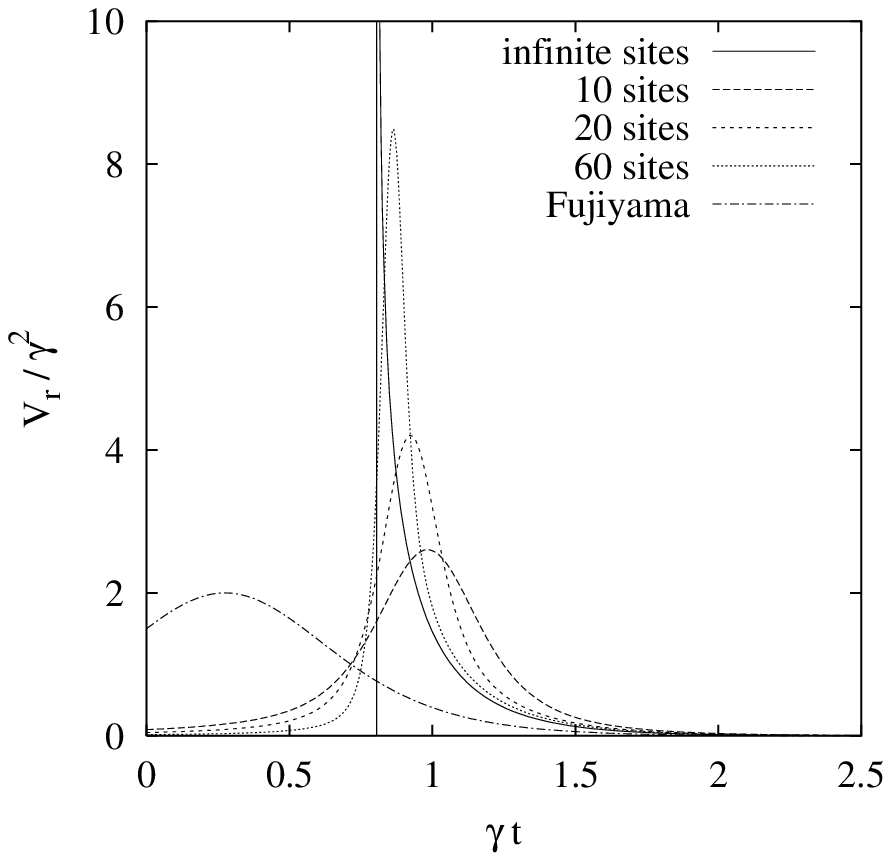}}
\caption{Time evolution of the equidistribution of genotypes
  in the zero mutation-limit of the symmetric fitness model for different
  sequence lengths.}
\label{time2}
\end{figure}

As in the equilibrium phase diagram, the dynamical transitions may 
be of first or second order. 
\begin{itemize}
\item
Second order transitions are located at
$\tilde{t} = \gamma t = 1$ for $\tilde{\gamma} \le 1/4$ and at 
$\tilde{t} = 1/\tilde{\gamma}_1$ for the transition from the
disordered phase to the PP phase. The transition from the PP phase to
the ACGT phase is second order above $\tilde{\gamma}_1 \approx 1.9009$
and implicitly given through $2\tilde{t}_c = 1 +
\exp[2\tilde{\gamma}_1(\tilde{t}_c - 1)]$. A similar second order 
transition (with a one-component order parameter) has also been
observed in the two-state model \cite{Wag,WBG}. 
\item
In an interval around the symmetry point $\gamma_1 = \gamma$, the
system possesses a first order transition (in the sense that there is a
finite jump in the magnetization). Note that, in contrast to the 
equilibrium
case, also the surplus and even the mean fitness are discontinous on
this line, giving rise to a rather pronounced threshold effect in the
evolution dynamics (cf.\ the solid line in Fig.~\ref{time2} 
for $\tilde{\gamma} = 1$). 
\end{itemize}
As for the equilibrium values, we also consider the effect of finite
sequence lengths on the time evolution. Again, calculations are
performed by direct diagonalization of the symmetric fitness model 
($\tilde{\gamma} = 1$). Fig.~\ref{time2} shows how the jump 
discontinouity in the mean fitness (internal energy) and the 
delta function singularity in the variance of the fitness (specific heat)
are approached by the finite systems. A threshold phenomenon is absent 
in the time evolution of the Fujiyama model which is also shown 
in Fig.~\ref{time2}.

%\begin{figure}[ht]
%\centerline{\epsfxsize=60mm \epsfysize=50mm \epsfbox{zvarfit1.ps} 
%\epsfxsize=60mm  \epsfysize=50mm \epsfbox{zvarfit2.ps}}
%\caption{Time evolution of the symetric fitness model for different
%  starting configurations.}
%\label{time1}
%\end{figure}

\begin{figure}[ht]
\centerline{\epsfxsize=130mm \epsfysize=45mm \epsfbox{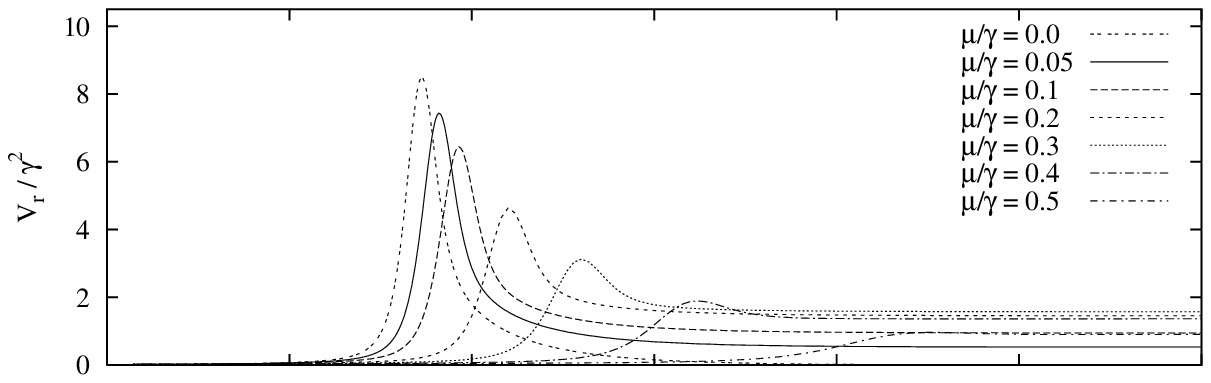}}
\vspace*{-12mm}
\centerline{\epsfxsize=130mm  \epsfysize=90mm \epsfbox{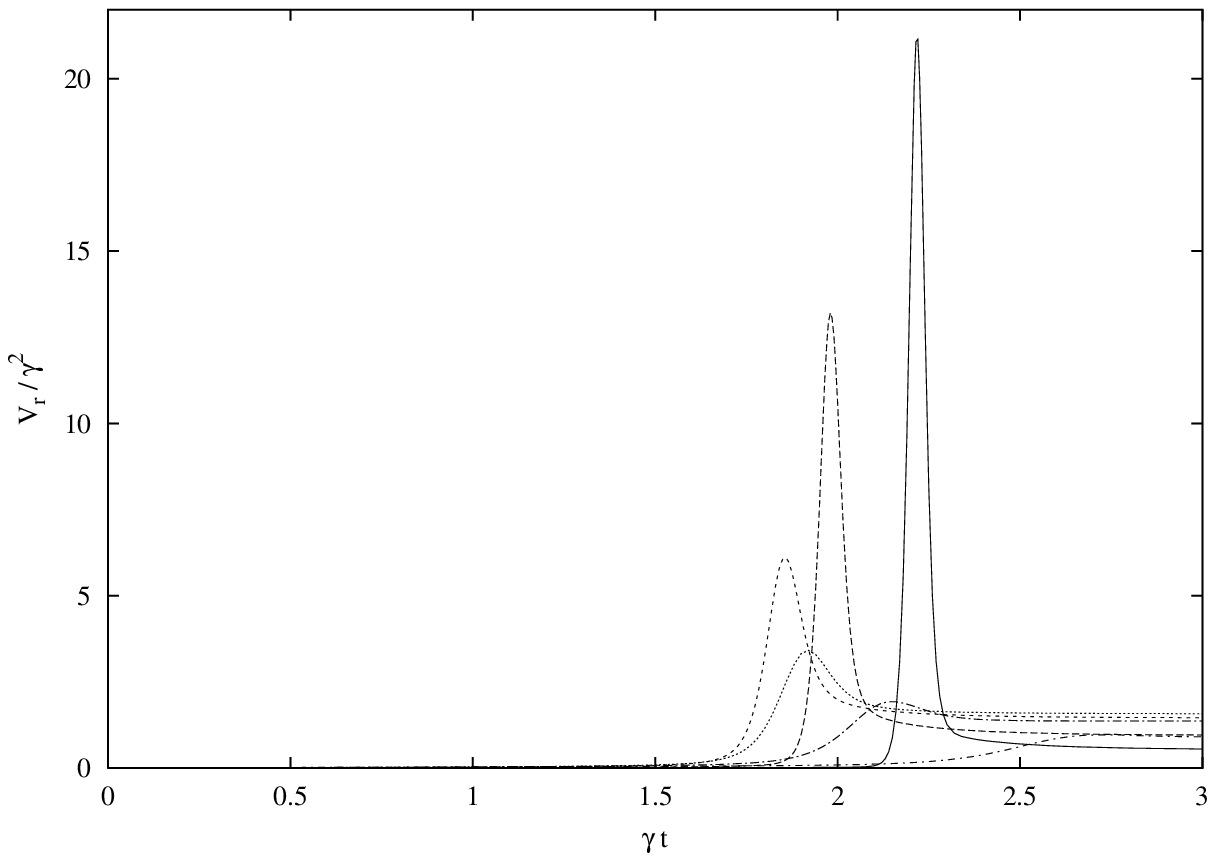}}
\caption{Time evolution of the variance of the fitness in the symmetric 
fitness model with sequence length $N=60$. Results are shown for
varying mutation rates and two different starting configurations.}
\label{time3}
\end{figure}

\paragraph{Finite mutation rates and different starting configurations}
In a last step, we now discuss the influence of the mutation rate and
the starting configuration on the evolution dynamics. Consider first the
time evolution of the equilibrium distribution of genotypes
$4^{-N}|\Omega\rangle$. Although no analytical results are available here, 
we may give the following intuitive argument that there is a phase 
transition at finite $t = t_c$ for any mutation rate below the
critical equilibrium mutation rate $\mu_c$: Since mutation alone tries to
keep the population in the equilibrium distribution, the evolution
dynamics will be slowed down by mutation for small $t$. In particular,
mean fitness and surplus will remain zero on a finite interval at
least up to the threshold value of the corresponding model with
vanishing mutation. On the other hand, the limiting values of $w$ and
$u$ are finite for $\mu < \mu_c$, giving rise to a non-analytical
point of $w(t)$ and $u(t)$ at some finite $t = t_c$. As shown in the
upper graph of Fig.~\ref{time3}, this behaviour is clearly visible in 
numerical results for finite sequence sizes. 

In order to contrast the time evolution of the unstructured population with
an equidistribution of genotypes as starting configuration, we have
also performed calculations for the opposite case of a population with
initially homogeneous phenotypes. Here, at $t=0$, any "individual" 
in the population has the same value $s_i = 0$ for the three surplus 
components. The result (for finite sequence length $N=60$) is shown 
in the lower viewgraph of Fig.~\ref{time3}. As for the
equidistribution of
genotypes, there is a clear threshold effect in the time evolution for 
any finite value $0<\mu<\mu_c$ of the mutation rate. The transition 
appears to be particularly sharp for small mutation rates. In contrast to the
unstructured case, the critical waiting time $t_c$ for the transition
is no longer monotonously increasing with the mutation rate $\mu$, but
is separated in two regimes: For mutation rates near the equilibrium 
threshold value $\mu_c$, the situation is similar to the unstructured
case: Here, single mutants with higher fitness appear in the
population after a short while. Due to the continuing mutation
pressure, however, a certain time is needed for these fitter
individuals to grow to a finite proportion and to dominate the mean
values in the infinite population. For small $\mu$, on the other hand, 
the critical waiting time $t_c$ is dominated by the time needed for
mutation to explore the configuration space and to generate 
individuals with higher fitness at a sufficient rate.

\section{Discussion}

When in \cite{BBW} a class of models for sequence space evolution was 
introduced, using the framework of Ising quantum chains, the calculations 
started with four major simplifications of the biological situation. 
These are the consideration of a two-state model, the assumption of an
infinite sequence length, the use of simplistic fitness landscapes,
and the restriction on infinite population sizes. In this paper, we
have looked at the first two of these simplifying assumptions. 
Finally, an extended discussion of the evolution dynamics of these 
models has also been presented. In the following paragraphs, we give 
a summary of our findings and an outlook on the remaining open problems.

\paragraph{Two-state versus four-state models.}
The main concern of this contribution is the generalization of the
modelling framework, introduced in \cite{BBW}, to four states
(corresponding to the four nucleotides) on each site. The
generalization presented makes use of the $C_2 \times C_2$ symmetry
inherent in the {\em Kimura 3 ST} mutation scheme. On the `physical 
side' this leads to a model of two coupled Ising quantum chains
(rather than to a four-state Potts model). Compared with the two-state
model, the extension can be thought of as consisting of two steps. In
a first step, we represent the four states on each site by the spin
values of two spins in decoupled chains. Note that already in this
simplified model three phases occur in the phase diagram since the
transition lines of the two decoupled chains will not in general 
coincide. The second step consists of the introduction of 
a more realistic mutation scheme which also changes the configuration 
space topology and the corresponding use of a refined fitness landscape.
Both these extensions lead to a coupling of the chains, and an even
richer phase space structure is found, including first-order transitions.
As may be seen from the introduction of a small linear field term into the
fitness function in subsection 4.2, this change of the transition to
first order leads to an increased robustness of the threshold
phenomena with respect to symmetry-breaking perturbations.

\paragraph{Finite sequence length.}
Typical sequence lengths of enzymes or viruses are of the order $10^3$
-- $10^4$. While these numbers are certainly far off the typical sizes of
macroscopic systems in physics, they are, in principle, large enough
to successfully supress $1/N$-corrections. However, especially models 
with simple fitness landscapes describe -- at best -- the evolution 
dynamics in a very restricted configuration space of particularly 
`important' sites, disregarding neutral or altogether lethal
mutations. In view of this fact, consideration of finite sequence 
lengths is indispensible and calculations in the thermodynamic
limit even seem to be questionable at first sight. In order to clarify
the usefulness of infinite-size methods in this context, we performed
a number of numerical calculations for finite sequence lengths. The
results are quite encouraging. As shown in subsection \ref{fs}, the 
characteristic properties of the thermodynamic limit are well visible 
even for tiny sequence sizes, such as $N = 10$, and the approximation 
is already quantitatively reasonable for sequences of length $60$.   

\paragraph{The fitness landscape.}
The construction of a tractable fitness landscape which nevertheless 
comprises the relevant biology is certainly the major task for all
these models. In this contribution, in order to obtain at least some 
analytical 
results, we have chosen a fitness function from the smooth end of the 
landscape zoo. Due to its permutation invariance, the quadratic
fitness function effectively disregards any local variance in
the interaction between sites, but only considers the average epistatic
effect. As such, it is in many respects certainly no more than a 
toy-model for evolution. However, the assumption of permutation
invariance of the sites is quite common in evolutionary biology and
comprises a large number of standard models for evolution, such as the
quadratic optimum model or Eigen's original sharply peaked landscape.
The results show that the essential structure responsible
for characteristic effects such as the error threshold is already
contained in this simplified framework and may 
also serve as a reference for future work on fitness functions
with increased ruggedness, such as the NK-landscape hierarchy \cite{KL}.
Here, we expect the results for the quadratic fitness model to be
qualitatively stable at least under certain forms of mild ruggedness,
such as the introduction of site-randomness in the fields and
interactions \cite{DK}. Pronounced changes, on the other hand, should
be expected when spin-glass effects come into play.

\paragraph{Finite population size.}
In going from the deterministic limit to the evolution of finite 
populations, the ordinary differential equation (\ref{paramuse}) has 
to be replaced by the master equation of a stochastic process which is
no longer covered by the theoretical framework presented in this
article. Due to the complexity of the stochastic equations, analytical results
seem to be out of reach at present for all but the simplest selection
schemes. Monte-Carlo simulations, however, should be possible and
could considerably add to theoretical insight here.

Although the general picture of the deterministic case should persist
at least for sufficiently large populations, the study of finite
population effects is certainly of importance.
For related models, such as the quasispecies model with the
{\em single peaked} landscape, it is has been found \cite{NS}
that the deterministic 
results can be interpreted as the time averages of the stochastic
process for mutation rates outside a certain interval around an error 
transition. Directly at the threshold, however, large fluctuations and 
a jump in the long-time averages appear in the stochastic system at a critical
mutation rate which seems to be lower by an amount roughly
proportional to $1/\sqrt{N}$ in comparison with the deterministic case.
Mainly because of these expected finite population effects we have
restricted discussions in this article entirely to the phase space 
structure of the models and the order of the phase transitions. Any further 
details of the transitions, even critical exponents, will presumably 
never be visible in real biological systems and thus seem to be
of limited relevance in this context.

Let us finally remark that, although biological populations are
certainly finite, the consideration of the infinite population limit
is not (only) a technical necessity, but also of direct importance for the 
study of the error threshold. That is so because this effect, in distinction
to the phenomenon of Muller's ratchet, is {\em by definition} not due to 
genetic drift, but solely due to the form of the fitness function. It
has thus always to be shown that the threshold effect persists even
for infinitly large population sizes.

\paragraph{Error threshold behaviour.} 

Since there are more than one and sometimes conflicting definitions of
the error threshold in literature (cf.\ the discussion in \cite{BG}), 
let us start this paragraph with a few clarifying remarks. In this 
article,
following \cite{BG}, we use the notion of the error threshold as
equivalent to phase transitions. As such, a clear-cut mathematical
definition (as non-analytical points in the mean fitness) is possible
only in the infinite sites (or thermodynamic) limit. However, since
the thermodynamic limit can be considered as an excellent
approximation already for rather small systems, the infinite system
property gives a valid explanation for prominent features which are
observable for finite sequences as well. In our study, we have always
considered sequences of a fixed length and have treated the mutation
rate per site as the variable driving the transition. In comparing
systems of different length, we have scaled the variables such that a
well-defined limit is approached as $N \to \infty$. In particular, the 
`critical' mutation rate per site in a finite system quickly converges
to the limiting value $\tilde{\mu}_c$. 
Originally, the threshold has been viewed as a limitating factor on 
the sequence length \cite{E}. This, however, should not be confusing: 
We switch to this latter picture simply by letting the reduced 
mutation rate depend linearly on the sequence length, 
$\tilde{\mu} \sim N$, and obtain a critical length 
$N_c \sim \tilde{\mu}_c$ (for sufficiently large sequences).

Our results on the error threshold phenomenon fit previous ones for
the two-state case and related models in that negative epistasis is
needed to observe a transition (cf.\ \cite{W,BG}).
Contrary to the two-state case, the threshold corresponds to a
first-order transition for certain parameter ranges and persists for
a sufficiently small linear part in the fitness function. Both, the
equilibrium and the dynamical phase diagram of the 
transition-transversion model (with $\alpha_i = 0$), 
possess two ordered phases characterized by non-zero values of one or 
all three components of the surplus order-parameter and the disordered 
phase with zero surplus where selection ceases to operate. The
threshold effect appears to be especially sharp in the evolution
dynamics, where a jump in the mean surplus and fitness and a delta
singularity in the variance of fitness occurs.

Besides the threshold effect, however, other properties of
mutation-selection models may be studied within the framework
presented. After all, exclusive concentration on phase
transitions is perhaps too much a physicist's point of view on these
systems. The relations between surplus, mutation rate and the variance of
fitness (\ref{zeit}), (\ref{variance}), for example, are valid for the entire
time evolution and arbitrary mutation rates. Depending on the fitness 
function applied, they may give rise to characteristic features also 
far off the transition point. This is particularly explicit for the
equilibrium variance of fitness which runs through a pronounced
maximum for fitness functions with negative epistasis at a mutation
rate much smaller than the threshold value.

\section*{Acknowledgments}

It is our pleasure to thank Ellen Baake and Oliver Redner for numerous
discussions and comments on the manuscript. Financial support from the
German Science Foundation (DFG) is gratefully acknowledged.

%\appendix{Threshold criterion for the symmetric model}

%In the 

%\begin{equation}
%f(\bm{\sigma}) := 3N \sum_{n=0}^\infty \left(\frac{c_n^{}}{n} 
%s^n(\bm{\sigma}) \right) \;;\quad  s_1 = s_2 = s_3 = s \;.     
%\end{equation}

%\begin{equation}
%hkgjkgh
%\end{equation}

\end{document}